\documentstyle[12pt,epsfig]{article}
   \font\tenmsb=msbm10 scaled\magstep 1
   \font\sevenmsb=msbm7 scaled \magstep 1
   \font\faivemsb=msbm5 scaled \magstep 1
\newfam\msbfam
      \textfont\msbfam=\tenmsb
      \scriptfont\msbfam=\sevenmsb
      \scriptscriptfont\msbfam=\faivemsb

\font\tengothic=eufm10 scaled\magstep 1
\font\sevengothic=eufm7 scaled\magstep 1
\newfam\gothicfam
      \textfont\gothicfam=\tengothic
      \scriptfont\gothicfam=\sevengothic

\newcommand{\be}{\begin{equation}}
\newcommand{\ee}{\end{equation}}
\newcommand{\Dlt}{\Delta}

\newcommand{\prt}{\partial}

\newcommand{\bt}{\beta}
\newcommand{\vp}{\varphi}

\newcommand{\al}{\alpha}
\newcommand{\ra}{\rightarrow}
\newcommand{\sgm}{\sigma}

\newcommand{\om}{\omega}

\newcommand{\lbd}{\lambda}

\begin{document}

\begin{center}

{\Large{\bf Nonlinear Dynamical Model of Regime Switching Between
Conventions and Business Cycles}  \\ [5mm]
V.I. Yukalov$^{1,2}$, D. Sornette$^1$, and E.P. Yukalova$^3$} \\ [3mm]

{\it $^1$Department of Management, Technology and Economics, \\
ETH Z\"urich, Z\"urich CH-8032, Switzerland, 
\\ [2mm] 

$^2$Bogolubov Laboratory of Theoretical Physics, \\
Joint Institute for Nuclear Research, Dubna 141980, Russia, \\ [2mm]

$^3$Laboratory of Information Technologies, \\
Joint Institute for Nuclear Research, Dubna 141980, Russia}

\end{center}

\vskip 2cm


\vskip 2cm


\newpage

\begin{abstract}

We introduce and study a non-equilibrium continuous-time dynamical 
model of the price of a single asset traded by a population of 
heterogeneous interacting agents in the presence of uncertainty and 
regulatory constraints. The model takes into account (i) the price 
formation delay between decision and investment by the second-order 
nature of the dynamical equations, (ii) the linear and nonlinear 
mean-reversal or their contrarian in the form of speculative price 
trading, (iii) market friction, (iv) uncertainty in the fundamental 
value which controls the amplitude of mispricing, (v) nonlinear 
speculative momentum effects and (vi) market regulations that may 
limit large mispricing drifts.  We find markets with coexisting 
equilibrium, conventions and business cycles, which depend on (a) 
the relative strength of value-investing versus momentum-investing, 
(b) the level of uncertainty on the fundamental value and (c) the 
degree of market regulation. The stochastic dynamics is characterized 
by nonlinear geometric random walk-like processes with spontaneous 
regime shifts between different conventions or business cycles. This 
model provides a natural dynamical framework to model regime shifts 
between different market phases that may result from the interplay 
between the effects (i-vi).

\end{abstract}

\newpage

\section{Introduction}

The concept of equilibrium has been central in economics for decades.
However, it is generally understood that equilibrium is just the first
approximation to the real market, valid on average and for long time
scales under certain idealized conditions. While equilibrium models 
have been quite useful in economics, real markets exhibit signatures 
of non-equilibrium, best exemplified during speculative bubbles (White,
1996) by the many documented instances of herding behavior (Shefrin,
2000). This non-equilibrium may lead to divergence from fundamental
value, anomalously large volatility (Shiller, 1989) and sometimes
unstable abrupt price variations in rallies and crashes (Sornette,
2003). The behavior of the market is said to be nonlinear as the price
set can be different depending for instance on the number of interacting
agents. The non-equilibrium nature of markets is especially pronounced in 
the sharp price variations occurring at booms and crashes, which are also 
accompanied by wild price fluctuations (Sornette, 1998; Sornette and 
Johansen, 2001; Sornette, 2003; Broekstra et al., 2005). These regimes 
require the use of non-equilibrium dynamical models (Farmer, 2002). 

Here, we propose and study a non-equilibrium continuous-time dynamical
model of the price of a single asset traded by a population of 
heterogeneous interacting agents in the presence of uncertainty and 
regulatory constraints. The model is defined by the equations (\ref{16}, 
\ref{18}) derived below, based on symmetry and structural constraints. 
Such a constructive scheme has a long tradition in the mathematical 
theory of dynamical systems and in the statistical physics of complex 
assemblies of elements (atoms, molecules, spins, and so on). This 
tradition is based on both the successes of this approach and on the 
many verifications that the structure of the obtained equations can 
be justified and derived from a full calculation involving all possible
interactions between the microscopic elements, when this is feasible 
(Wilson, 1979). Such a tradition is non-existent in economic theory 
but we think it is inspiring to attempt to use and adapt in the goal 
of obtaining new insights. Here, we acknowledge the huge difficulty 
in first describing correctly the decision making process of individual 
investors, and second in developing a first-principled approach of the 
aggregate behavior. The symmetry- and structural-based constructive 
method that we propose bypasses these difficulties by guiding us towards 
what should be the generic structure of the macroscopic processes that 
survive and capture the transition from the microscopic worlds of many 
interacting agents to the macroscopic collective market.

The model, that we obtain, has two specific characteristics. 
First, in contrast with most models, it is second-order in its time 
derivative to account for the temporal delays between agents' actions 
and the formation of price. Second, it emphasizes the importance of 
taking into account strong nonlinear behaviors in a "non-perturbative" 
way in order not to rule out possible large deviations from equilibrium 
and to describe them correctly. For this, we suggest a novel way of 
treating nonlinear terms, which yields nonpolynomial nonlinearities. 
The form of the latter automatically stabilizes the market dynamics, 
without the necessity of introducing artificial constraints. A main goal 
of the model is to understand the conditions under which different price 
regimes may result from the competition between mean-reversal to 
equilibrium and herding. This competition is controlled by (i) the degree 
of uncertainty in the fundamental value of the asset and (ii) how market 
regulations may constrain large price variations.

The work most closely resembling ours is the adaptation to financial 
markets by Broekstra et al. (2005) of Weidlich's (2000) extended model 
of opinion formation and political phase transitions. This model is also
formulated as a nonlinear system of two ordinary differential equations
of two dynamical variables, hence it supports cycles as does our model
with second-order dynamics. The two interacting variables are the global 
unbalance between supply and demand for a share and the personal preference
or sentiment of agents. As our model, Weidlich's (2000) model exhibits fixed 
points, cycles and exponentially growing unstable regimes. Based on a 
micro-foundation of transition rates depending exponentially on the number 
of investors in buy or sell positions, it however lacks the "convention" 
fixed points that we unravel and the nonlinear blow-up regimes, which have 
been found to characterize well explosive bubbles and crashes (Sornette, 
2003). 

Our main results are the derivation of a new nonlinear and nonequilibrium 
model of financial markets, a detailed investigation of its dynamics, 
classification of all possible dynamical phase portraits corresponding 
to qualitatively different types of markets, and the study of the role 
of stochasticity, which is shown to be a triggering mechanism for the 
transition of the price trajectories between different dynamic regimes.

The organization of the paper is as follows. We present in section 2 
the derivation of the dynamical equations for the asset price dynamics. 
Section 3 gives the classification of the limiting and degenerate cases 
of the dynamics, which serve as pivots to understand all the different 
regimes. Section 4 derives the full classification of the different market 
regimes. This dynamics turns out to be unusually rich, demonstrating several 
nontrivial bifurcations leading to coexisting prices (equilibrium on the 
fundamental value and "conventions") as well as cycles. The main 
dynamical regimes of the market evolution are illustrated by the related 
phase portraits and by direct representations of the time dynamics of the 
price. The different regimes depend on (a) the relative strength of 
value-investing versus momentum-investing, (b) the level of uncertainty 
on the fundamental value and (c) the degree of market regulation.

\section{Derivation of Price Dynamics}

\subsection{General structure}

Market dynamics is usually described by the time dependence of an 
asset price $p(t)$ or the log-price $\log p(t)$, where the logarithm 
is understood as the natural logarithm. After Bachelier (1900), one has 
standardly considered first-order stochastic differential equations. 
But Farmer (2002) emphasized that, for the correct description of 
nonequilibrium markets, one has to deal with second-order dynamical 
equations, which take into account temporal delays between agent actions 
and the formation of price. In this spirit, second-order differential 
equations for the log-price were considered by Bouchaud and Cont (1998) 
and by Grassia (2000).

Consider an asset with market price $p(t)$ at time $t$. The anticipated
future discount factors and expected flow of dividends provide in 
principle the fundamental value $p_f(t)$ of the asset. But, $p_f(t)$ is
not directly observable, so that we define the mispricing variable
as the difference between the logarithm of the market price
and the logarithm of the fundamental price (Ide and Sornette, 2002)
\be
\label{3}
x \equiv \log p - \log p_f \; .
\ee
We assume that $x$ follows the following process
\be 
\label{4}
dx = y dt + \sigma dW \; .
\ee
In equilibrium markets, the drift term $y$ in (\ref{4}) is determined
(He and Leland, 1993) from the von Neumann-Morgenstern utility function
of the agents (von Neumann and Morgenstern, 2004). For non-equilibrium
markets, we consider that the drift $y$ is determined as a function of
state variables. In the equilibrium-based models of Cecchetti, Lam, and
Mark (1990, 1993) and Bonomo and Garcia (1994a, b, 1996) for instance,
the state variable governs the evolution of the fundamentals of
consumption and dividends of the economy. Here, we add the new
ingredient that these fundamentals of the economy are not exogeneous. 
We account for the possibility that the state variables are themselves
depending endogeneously on the mispricing $x$ and drift term $y$. This
structure implies a feedback mechanism capturing the reflexivity and
self-consistency of financial market resulting from the collective
organization of investors. This represents what G. Soros refers to as
``reflexivity'', that is, ``actors observing their own deeds,'' such
that ``market participants are trying to discount a future that is
itself shaped by market expectations.'' Mathematically, we thus
postulate the following process for the drift term
\be
\label{6}
dy = f(x,y,t) dt + \sigma' dW' \; ,
\ee
$W$ and $W'$ are two standard (possibly correlated) Wiener processes.
The right-hand side of (\ref{6})
represents the overall contributions of all market ``forces,'' that 
may influence the market price to drift away from the fundamental 
price. Generally, this ``force'' depends on time and may also contain 
random terms. Note that (\ref{4}) together with (\ref{6}) makes the 
dynamics of the mispricing effectively of second-order. This accounts 
for the delay between the decision and impact of agents on the price 
forming process. 

The structure of $f(x,y,t)$ in (\ref{6}) should give the possibility 
to describe strongly nonequilibrium markets, with large price variations, 
such as occurring during financial bubbles and crashes. In this spirit, 
several nonlinear equations have been studied aiming at characterizing 
the first-order dynamics of price during financial bubbles (Sornette and 
Andersen, 2002; Andersen and Sornette, 2004) and the second-order dynamics 
of price or log-price during stock market booms and crashes (Bouchaud and 
Cont, 1998; Pandey and Stauffer, 2000; Ide and Sornette, 2002). A system 
of equations for price, which can be reduced to a third-order differential 
equation has also been suggested (Thurner, 20001).

\subsection{Symmetry, Taylor expansion and resummation method}

Equations (\ref{4}) and (\ref{6}) describe the general market evolution. 
To specify the market "force", let us consider the case, when there are 
no external influences in addition to those impacting directly the 
fundamental value $p_f(t)$, so that the force is defined by the market 
structure itself, and $f(x,y)$ can be considered independent of time. 
As usual, we assume that the driving market force $f(x,y)$ is an analytic 
function of its arguments, hence, it can be represented as a Taylor 
expansion
\be
\label{7}
f(x,y) = \sum_{m,n} c_{mn} x^m y^n \; .
\ee

Our aim is to model the market force in such a form that it would be, 
on the one hand, sufficiently simple and, on the other hand, catching 
the basic properties of the market structure as well as the agent's 
collective characteristics and style. To this end, we notice that the 
influence of the cross-terms $x^m y^n$ on the dynamics are analogous to 
the sum of terms $x^{m+n}$ and $y^{m+n}$. Therefore, it is possible to 
neglect the cross-terms, accordingly renormalizing the coefficients of 
the terms $x^m$ and $y^n$. This is equivalent to saying that the force 
is additive,
\be
\label{8}
f(x,y) = f_1(x) + f_2(y)
\ee
being the sum of the terms
\be
\label{9}
f_1(x) = \sum_{n=0}^\infty a_n x^n \; , \qquad 
f_2(y) = \sum_{n=0}^\infty b_n y^n \; .
\ee
     
Another simplification that one often uses is the assumption that 
the market is symmetric, in the sense that there is no drastic 
difference between rising and falling prices (Farmer, 2002). Of 
course, this is not exactly true for real markets, but can be 
accepted as a reasonable approximation. Here, we assume a weaker 
symmetry, that is, the dynamics is the same for upward as for 
downward mispricing $x$. This does not prevent the observable asset 
price $p(t)$ to present asymmetric structures, which can then be 
traced to the asymmetric behavior of the fundamental price $p_f(t)$. 
For instance, the so-called leverage effect is not excluded in our 
model. Indeed, in its initial formulation (Black, 1976), it expresses 
an inverse relationship between the level of equity prices and the 
instanteneous conditional volatility: a drop in the price of the 
stock increases the debt-to-equity ratio and therefore the risk of 
the firm which translates into a higher volatility of the stock. This 
mechanism acts at the level of the fundamental price $p_f(t)$ and can 
thus be straightforwardly incorporated in a suitable dynamics of $p_f$, 
whose specification is not needed here in our first effort emphasizing 
the different classes of mispricing dynamics. More complex forms of
leverage effects which could appear on mispricing (Figlewski and 
Wang, 2000) are excluded in our model.

Within our formulation, the symmetry of the mispricing implies that 
the evolution equations (\ref{4}) and (\ref{6}) should be invariant 
under the inversion transformation
\be
\label{10}
x \; \ra \; -x \; \qquad y \; \ra \; -y \; \qquad 
W \; \ra \; -W \; .
\ee
This requires that the market force be asymmetric, such that
\be
\label{11}
f(-x,-y) = - f(x,y) \; .
\ee
Thence, in the Taylor expansions for $f_1(x)$ and $f_2(y)$, 
only the odd terms remain, while there are no even terms, since 
$a_{2n}=b_{2n}=0$. 

Separating the linear and nonlinear terms in the force, we may 
write
\be
\label{12}
f_1(x) = a_1 x + a_3 x^3 \vp_1(x) \; , \qquad 
f_2(y) = b_1 y + b_3 y^3 \vp_2(y) \; ,
\ee
where
$$
\vp_1(x) \equiv 1 + \frac{a_5}{a_3}\; x^2 + 
\frac{a_7}{a_3}\; x^4 + \ldots \; ,
$$
$$
\vp_2(y) \equiv 1 + \frac{b_5}{b_3}\; y^2 + 
\frac{b_7}{b_3}\; y^4 + \ldots \; .
$$

Nonlinearities are commonly introduced into the market dynamics by 
assuming that the effective force driving the market can be expanded 
in Taylor series and by limiting these series by several first terms 
(Farmer, 2002). The general weak point of nonlinear equations with 
polynomial nonlinearities is the appearance of unstable solutions, 
when either the market price, or the price rate, or both become 
divergent, often even at a finite moment of time (Ide and Sornette, 
2002). Actual divergences in markets are, of course, unrealistic. 
Therefore, as soon as there arise singular solutions, this means 
that the corresponding market is absolutely unstable and cannot exist. 
Alternatively, one could impose additional constraints regularizing the 
admissible price variations (Grassia, 2000; Farmer, 2002). However, 
externally imposed constraints practically kill the influence of 
nonlinear terms and make the market dynamics rather poor, not allowing 
for large price fluctuations typical around booms and crashes. The 
latter phenomena are principally collective and strongly nonlinear, 
somewhat similar to phase transitions and critical phenomena, where the 
characteristic nonlinearities are to be appropriately treated (Yukalov 
and Shumovsky, 1990; 
Sornette, 2006).

Another issue with the standard way of dropping higher-order terms, 
leading to $\vp_1(x) = 1$ and $\vp_2(x) = 1$, is that this would assume 
that there are no large or fast price variations, so that the market is 
close to equilibrium and therefore the values $x$ and $y$ can be treated 
as small. But as we wish not to rule out the possibility of describing 
strongly out-of-equilibrium markets, where the mispricing can be large 
and its change fast, it would be principally incorrect to omit the 
higher-order terms in the force. We conclude that {\it all} terms in 
$\vp_1(x)$ and $\vp_2(y)$ must be kept for correctly describing 
non-equilibrium markets.

However, keeping an infinite number of nonlinear terms would make the 
problem untreatable. At the same time, limiting the divergent series 
by a finite number of terms is principally wrong as we said earlier. 
The way out of this conundrum is to find an effective way to perform 
the sum in the infinite series by means of a resummation procedure. For 
this purpose, it is convenient to employ the self-similar approximation 
theory previously developed in (Yukalov, 1990, 1991, 1992; Yukalov and 
Yukalova, 1993, 1996, 1999, 2002; Gluzman and Yukalov, 1997). The variant 
of this theory, involving the self-similar exponential approximants 
(Yukalov and Gluzman, 1997, 1998) is the most appropriate for describing 
market processes (Gluzman and Yukalov, 1998; Andersen et al., 2000; 
Yukalov, 2000, 2001). By means of the self-similar exponential 
approximants, the series for $\vp_1(x)$ and $\vp_2(y)$ can be 
transformed into
\be
\label{13}
\vp_1(x) = \exp\left ( \frac{a_5}{a_3}\; x^2 \exp \left ( 
\frac{a_7}{a_5} \; x^2 \ldots \right ) \right ) \; ,
\ee
\be
\label{14}
\vp_2(y) = \exp\left ( \frac{b_5}{b_3}\; y^2 \exp \left ( 
\frac{b_7}{b_5} \; y^2 \ldots \right ) \right ) \; .
\ee
For what follows, in order not to overcomplicate the problem, it is 
sufficient to use the first-order exponential approximants
\be
\label{15}
\vp_1(x) = \exp\left ( -\; \frac{x^2}{\mu^2} \right ) \; , \qquad 
\vp_2(y) = \exp\left ( -\; \frac{y^2}{\lbd^2} \right ) \; ,
\ee
in which
$$
\mu^2 \equiv -\; \frac{a_3}{a_5}\; , \qquad 
\lbd^2 \equiv -\; \frac{b_3}{b_5}\; ,
$$
where the coefficients $a_3$ and $a_5$ (respectively $b_3$ and $b_5$) 
have opposite signs.

In this way, starting from equation (\ref{4}), we obtain the 
following system of equations (\ref{16},\ref{18}) describing the 
evolution of non-equilibrium market mispricing
\be
\label{16}
dx = y dt + \sigma dW \; ,
\ee
Equation (\ref{6}) for the 
drift takes the form
\be
\label{18}
\frac{dy}{dt} = \al x + \bt y + A x^3 \exp \left ( -\; 
\frac{x^2}{\mu^2} \right ) + B y^3 \exp \left ( -\; 
\frac{y^2}{\lbd^2} \right )  \; ,
\ee
in which, for convenience, we use the notation
$$
\al \equiv a_1 \; , \qquad \bt \equiv b_1 \; , \qquad 
A \equiv a_3 \; , \qquad B \equiv b_3 \; .
$$
In expression (\ref{18}), we have taken $\sigma'$ equal to zero
to emphasize the deterministic feedback processes controlling
the dynamics of the drift of the mispricing.

The equations (\ref{16},\ref{18})  constitute our basis for 
describing at a coarse-grained level the price dynamics resulting 
from the aggregate action of investors. The equations (\ref{16}) 
and (\ref{18}) have been introduced phenomenologically by using only 
very general symmetry and structural considerations, following the 
strategy of model constructions developed in the mathematics of 
dynamical systems described by ordinary differential equations (Arnold, 
1978). The strength of our derivation lies in capturing the main 
ingredients of the collective aggregate behavior of the asset price. 
The use of symmetry and structural reasoning allows one to bypass the 
need for the exceedingly difficult description of the dynamical decision 
making processes and interactions between heterogeneous agents. In this 
way, the theory is less prone to mispecification error, in the sense 
that that the global collective description can be thought of as more 
robust than a micro-foundation which is very susceptible to model 
errors. The weakness of our approach is that we do not use a 
microeconomic agent-based model where consumption preferences or 
utility functions appear explicitly. However, this is not necessarily 
a handicap as the aggregate decision making processes lead to a 
dynamical evolution equation, and it is the later that we attempt to 
capture in a robust way. The parameters of the theory are not determined 
quantitatively by some assumptions of agent preferences, but should be 
considered as general effective collective parameters whose meaning is 
elaborated below.

The equations (\ref{16},\ref{18})  provide us the opportunity to 
study what are the conditions for market efficiency, when and how does 
the price converge to equilibrium when starting from non-equilibrium 
initial conditions, what are other possible regimes which may result 
from the interplay between uncertainty, herding and market regulations?

\subsection{Economic and behavioral interpretation}

Before presenting a detailed study of equations (\ref{16}) and 
(\ref{18}), it is useful to extract the economic meaning of the 
parameters $\al,\bt,A,\mu,B,\lbd$ appearing in equation (\ref{18}). 
First, the dynamics of the mispricing variable $x=\log (p/p_f)$ is 
clearly of second-order in the time derivative, in agreement with 
Farmer's (2002) arguments. The form of the equations shows that 
this is a non-Hamiltonian system, with the implication that there 
is no conserved quantity such as a probability or energy, as can 
be expected for a driven out-of-equilibrium system. The terms in the 
right-hand side of (\ref{18}) express the different contributions of 
the market reactions due to deviations from equilibrium ($x \neq 0$ 
and $y \neq 0$), that impact the mispricing drift $y$. 

\begin{itemize}

\item 
The term $\alpha x$ expresses a mean reversal if $\alpha<0$ 
resulting from value investment styles. However, one should not 
exclude market phases in which $\alpha$ may be positive, reflecting 
a speculative behavior in which over-valuation (resp. under-valuation) 
triggers more demand (respectively less demand), pushing the price 
further up (resp. down) according to a positive feedback effect. This 
behavior is typical in market phases which are referred to as "bubbles".

\item 
The parameter $\beta$ of the second term of the r.h.s. of (\ref{18}) 
is taken as always negative to embody "market friction", i.e., the 
different costs and restraints, such as commissions and tax implications 
associated with transactions. We consider that market friction opposes 
the growth of the drift in mispricing, in order words, it tends to 
dampen large price variations.

\item 
The two nonlinear terms in (\ref{18}) characterize the collective behavior 
of traders, resulting from their interactions. The first nonlinear term 
$Ax^3\exp(-x^2/\mu^2)$ emphasizes or corrects the effect of the first term 
$\alpha x$ for large mispricing. The first part $A x^3$ adds on the linear 
term $\alpha x$ by stressing that a fraction of the population of investors 
may decide to correct mispricing ($A<0$) or speculate on further 
disequilibrium ($A>0$), only when the mispricing is sufficiently large. 
As argued in (Ide and Sornette, 2002), this may be due to the fact that 
investors have only an approximate (and perhaps even fuzzy) estimation 
of the fundamental value, since future dividends and discount rates can 
be assessed only with rather large uncertainty. As a consequence, these 
investors will only buy or sell when the mispricing is larger than their 
bracket of uncertainty. This creates a threshold-like behavior which is 
adequately represented by a nonlinear term such as $A x^3$. Ide and 
Sornette (2002) considered other powers $x^n$ with $n>1$. Here we stick 
to $n=3$ for simplicity.

\item 
The factor $\exp(-x^2/\mu^2)$ plays a significant role only for nonlinear 
collective speculative behavior characterized by $A>0$. Without this term 
and for $A>0$, the mispricing equations (\ref{16}) and (\ref{18}) lead to 
finite-time singularities similar to those analyzed by Ide and Sornette 
(2002), which cannot therefore describe a long-lived market, but only the 
price run-up during a transient explosive bubble or the price collapse 
during a crash. But in reality, the price cannot depart arbitrarily far 
away from the fundamental value, i.e., the mispricing $x$ cannot go to 
$\pm\infty$. If not the market forces, somehow regulations or 
implementation market constraints come into play, such as so-called 
circuit breakers. For instance, on October 27, 1997, circuit breakers 
caused the New York Stock Exchange to halt trading for the first time 
in history as the Dow Jones Industrial Average lost 554 points. 
Similarly, an upside mispricing cannot go on forever if investors have 
some information on the real value $p_f$. The saturation of the growth 
potential of mispricing is embodied in the parameter $\mu$, which can 
be thought of as a measure of uncertainty in the fundamental value. If 
the uncertainty is small, $\mu$ is small, and the mispricing is limited 
due to the collective convergence of informed investors. On the other hand, 
if the uncertainty is large, $\mu$ is large, and the mispricing can become 
important. The limit $\mu \to +\infty$ corresponds to the limit of absence 
of information on the fundamental value $p_f$.

\item 
The last term in Eq.~(\ref{18}) contains two contributions. The nonlinear 
factor $B y^3$ takes into account speculative trading behavior leading 
to a momentum effect, in which a detected trend is reinforced. We thus 
only consider that case $B>0$. Contrarian strategies would lead to $B<0$ 
but they are in general used by a minority of investors. Hence, the 
collective behavior is better reflected by the majority style which 
we assume to be $B>0$. The nonlinearity stresses the fact that, as argued 
by Ide and Sornette (2002), a trend is only detected when sufficiently 
strong. As a consequence, investors take action based on momentum strategies 
only when $y$ is sufficiently large, above some noise background, thus
leading to a kind of threshold effect well-captured by the third-order
nonlinearity. On the basis of similar arguments, Ide and Sornette (2002)
have also considered such a term $\sim y^m$ with $m>1$. Here, we stick
to $m=3$ for simplicity.

\item 
The last factor $\exp(-y^2/\lbd^2)$ embodies market regulation and 
$\lbd$ can be called the "liberatization" parameter. The larger it is, 
the more free are the changes of mispricing drift. In contrast, when the 
liberalization parameter is small, i.e., regulations are significant, 
large mispricing drifts are prevented by a rigid market.

\end{itemize}

To summarize the key points of the model of mispricing given by 
(\ref{16},\ref{18}), we take into account the price formation delay 
between decision and investment by the second-order nature of the 
dynamical equations and we account for (i) linear and nonlinear 
mean-reversal or their contrary in the form of speculative price 
trading, (ii) market friction, (iii) uncertainty in the fundamental 
value which controls the amplitude of mispricing, (iv) nonlinear 
speculative momentum effects and (v) market regulations that may 
act to limit large mispricing drifts.

\subsection{Nonlinear dynamics versus stochasticity}

The rest of the paper is devoted to a careful study of the system 
(\ref{16},\ref{18}). In fact, there are two elements to discuss. 
First, expression (\ref{18}) contains highly nonlinear deterministic 
terms which create, as we shall see, a rich phase diagram of different 
regimes. Second, the stochastic term in (\ref{16}), when injected in 
the dynamics, further enriches it. We can summarize the situation by 
saying that there is an interplay between the nonlinear deterministic 
dynamics with transient chaos and stochasticity. When omitting 
stochasticity, i.e., $\sigma dW$ to zero, we obtain, as described 
below, many interesting regimes with coexisting attractors (equilibrium, 
"conventions", nonlinear cycles). Then, reinjecting a not-to-large 
stochasticity $\sigma dW$  makes the dynamics follow the principal 
structures of the phase portrait obtained by neglecting it, leading to 
some distortion of the deterministic trajectories, but not destroying 
the existing attractors. Stochasticity mixes the basins of attraction 
of the different regimes and the basins of attraction are replaced by 
transient traps. A trajectory can live for quite a long time inside 
one of such trapping regions along a stochastic trajectory, but then 
jumps stochastically to another trapping region, where it again can 
spend a rather long time. This gives a dynamical description of regime 
shifts which can provide an underpinning of Markov Chain switching 
models (Hamilton, 1989; Hamilton and Raj, 2002). We shall illustrate 
this point at the end of the paper, when our classification of the 
different regimes from which the price can switch have been performed. 
Thus, our emphasis on the classification of the deterministic 
trajectories of (\ref{16}) and (\ref{18}) obtained when $\sigma=0$  
does not mean that we consider the market mispricing as deterministic. 
Rather, we emphasize the interplay between nonlinearity and stochasticity, 
which may lead to possible explanations of some of the most important 
stylized facts of financial time series, such as anomalous large 
volatility, long memory effect in volatility, bubbles and crashes and 
regime shifts. But for a deeper understanding of this interplay, it is 
necessary to first describe the structure of the deterministic attractions 
(Horsthemke and Lefever, 1984).

\section{Limiting and Degenerate Cases}

In order to better understand the role of the different terms in 
the evolution equations (\ref{16},\ref{18}), let us consider some 
limiting cases, when these equations degenerate to simple or known 
forms. In the study below, we assume that the noise term $\sigma dW$ 
can be neglected and the market friction parameter is  always negative, 
$\bt<0$.

The summary of the analysis, presented below in the different subsections, 
is provided in Table 1. For the convenience of notation, the "convention" 
fixed points $x_{2,3}^*$ are denoted as
\be
\label{19}
x_{2,3}^* = \pm s \; , \qquad s\equiv \sqrt{-\; \frac{\al}{A}} \; ,
\ee
so that $s$ corresponds to a positive mispricing convention and $-s$, to
negative mispricing convention. In the table, we show the conditions of
the global or local stability and the related stable fixed points. Local
stability is associated with the existence of a finite basin of attraction 
around the corresponding fixed point, that is, when this basin of 
attraction does not cover the whole phase plane. The results are obtained 
by means of the Lyapunov stability analysis as well as invoking numerical 
investigations of the related equations. The market friction coefficient 
$\bt$ is assumed to be always negative, $\bt<0$. And the nonlinear trend 
followers are supposed to be characterized by positive feedback, i.e., 
$B>0$.

In these limiting cases, a quick glance at Table 1 tells us that 
overregulation is always preferable to the absence of any regulation 
for both markets with no uncertainty as well as for markets with complete 
uncertainty on the fundamental value. Global stability can be achieved 
only in regulated markets, while the unregulated certain markets can be 
only locally stable.

\subsection{{\it no uncertainty} ($\mu\ra 0$) {\it over-regulated} 
($\lbd\ra 0$)}

In this case, the nonlinear terms responsible for the collective effects 
disappear. Specific collective behavior can not develop in the presence 
of strongly regulated markets in which the fundamental value is known 
information to all investors. Only the linear terms control the dynamics 
of Eq.~(\ref{18}):
\be
\label{20}
\frac{dy}{dt} = \al x + \bt y \; .
\ee
The only fixed point of the dynamics is the equilibrium point 
$\{ x^*=0,y^*=0\}$ corresponding to no mispricing. The eigenvalues of the 
corresponding Jacobian, which quantify the stability of this fixed point, 
are
\be
\label{21}
J^\pm = \frac{1}{2}\left ( \bt \pm \sqrt{\bt^2+4\al} \right ) \; .
\ee
Hence, if the traders have mean-reverting strategies $(\al<0)$, then 
$\{ 0,0\}$ is a stable fixed point and the market converges to its 
equilibrium, $x\ra 0$ and  $y\ra 0$. 

In contrast, if the investors exhibit a speculative behavior with positive 
feedback on the price, so that $\al>0$, then the dynamics becomes globally 
unstable and mispricing $x$ as well as mispricing drift $y$ diverge. The 
market can only be transient. Hence, fully informed irrational investors 
blow up the market. Only rational trading is compatible with the existence 
of strong regulation and full available public information. This is not 
a surprise.

\subsection{{\it maximum uncertainty} ($\mu\ra\infty$)  
{\it un-regulated} ($\lbd\ra\infty$)}

In this case, Eq. (\ref{10}) reduces to
\be
\label{22}
\frac{dy}{dt} = \al x + \bt y +Ax^3 +By^3\; .
\ee
There can exist three fixed points. The equilibrium fixed point $\{ 0,0\}$
is always present, with the related Jacobian eigenvalues
\be
\label{23}
J^\pm_1 = \frac{1}{2}\left ( \bt \pm \sqrt{\bt^2+4\al} \right ) \; .
\ee
The two additional fixed points
\be
\label{24}
x_{2,3}^* = \pm \sqrt{-\; \frac{\al}{A}}\; , \qquad 
y_{2,3}^* = 0\; ,
\ee
with the corresponding Jacobian eigenvalues
\be
\label{25}
J_{2,3}^\pm = \frac{1}{2}\left ( \bt \pm \sqrt{\bt^2-8\al} 
\right ) \; ,
\ee
exist only if $\alpha$ and $A$ have opposite signs.

For mean-reversing investors $(\al<0)$, only the equilibrium point 
$\{ 0,0\}$ is stable, the two other points being either saddle points 
for $A>0$, i.e. with one stable direction and one unstable direction, 
or do not exist at all for $A<0$.

For speculative traders $(\al>0)$, the point $\{ 0,0\}$ becomes a saddle
point. For $A>0$, there are no stable points and all trajectories diverge. 
When $\al>0$ and $A<0$, the two fixed points $\{ x_{2,3}^*,0\}$ are stable, 
and represent two conventions in the sense of Keynes and Orl\'ean (1994), 
(see also Boyer and Orl\'ean, 1992; Eymard-Duvernay et al., 2005), in which 
investors share a common belief which is self-realized with no real 
underpinning by a fundamental valuation. Note that $\{ x_{2}^*,0\}$ (resp. 
$\{ x_{3}^*,0\}$) corresponds to an overprice (resp. underpriced) convention.

When $\al$ crosses $0$, a bifurcation occurs, such that the rational 
behavior, with just one stable equilibrium point $\{ 0,0\}$, transforms 
into the speculative behavior characterized by the two stable convention 
points $\{ x_{2,3}^*,0\}$. This dynamical transition is analogous to the 
price symmetry breaking mechanism introduced by Sornette (2000), the 
difference being that, in our present case, the symmetry-breaking of 
$\{ 0,0\}$ into $\{ x_{2,3}^*,0\}$ occurs with respect to the reference 
fundamental price rather than to a vanishing price.

Both for the rational $(\al<0)$ and for the speculative $(\al>0)$ 
behaviors, the existing stable fixed points have only finite basins of 
attraction. This means that all trajectories, starting inside a given 
basin, tend to its corresponding fixed point. But any trajectory, with 
initial conditions outside the basins of attraction, diverges in a finite 
time $t_c$ as
\be
\label{26}
x\; \propto \; \sqrt{t_c - t}\; , \qquad 
|y| \; \propto \; (t_c-t)^{-1/2}\; ,
\ee
for $t\ra t_c-0$. This singular behavior occurs, as we said above,
due to the complete uncertainty ($\mu \to +\infty$) on the fundamental 
price and the complete lack of regulation ($\lambda \to +\infty$), which
remove any restriction or constraint on the mispricing. The market can be 
said to be only locally stable, but it is globally unstable.

In the limit where the linear terms can be neglected compared to the 
nonlinear terms, so that $\al\ra 0$ and $\bt\ra 0$, the dynamics reduces 
to the equation
\be
\label{27}
\frac{dy}{dt} = Ax^3 +By^3\; ,
\ee
which is a variant of the Ide-Sornette (2002) model. Then, the basins 
of attraction disappear completely. There are no stable fixed points, and
all trajectories diverge in finite time. The model (\ref{27}) is structurally 
unstable, since its dynamics sharply changes under infinitesimally small $\al$ 
and $\bt$. Its usefulness lies however in providing a structural description 
for the transient behaviors associated with explosive bubbles and their crashes 
(Ide and Sornette, 2002).

\subsection{{\it maximum uncertainty} ($\mu\ra\infty$) {\it over-regulated} 
($\lbd\ra 0$) \label{ngnlflvq}}

Equation (\ref{18}) becomes
\be
\label{28}
\frac{dy}{dt} = \al x + \bt y + Ax^3 \; .
\ee
In this limit where the nonlinear term $By^3\exp(-y^2/\lbd^2)$ is absent, 
the impact of trend followers vanishes (since over-regulation prevents any 
trend to appear). Mathematically, equation (\ref{28}) is known as the 
Duffing equation. 

In the case of mean-reversing traders $(\al<0)$, the equilibrium point 
$\{ 0,0\}$ is the unique fixed point. When speculative traders dominate 
$(\al>0)$, $\{ 0,0\}$ becomes a saddle, while two other stable fixed 
points $x_{2,3}^*=\pm\sqrt{-\al/A}$ appear, provided that $A<0$. When 
$A<0$, the dynamical system is globally stable, with one stable fixed 
point $\{ 0,0\}$ if $\al<0$ or with two stable fixed points $\{ x_{2,3},0\}$ 
if $\al>0$. When $A>0$, then there is one stable fixed point $\{ 0,0\}$ if 
$\al<0$, with a finite basin of attraction. For $\al>0$ and $A>0$, there are 
no stable fixed points and all trajectories diverge. Hence for $A>0$, the 
system is globally unstable, since there are trajectories, starting outside 
the basin of attraction (when it exists), which diverge. For speculative 
traders $(\al>0)$, the market bifurcates either to the overvalued convention 
$\{ x_2^* >0,0\}$ or to the unvalued convention $\{ x_3^* <0,0\}$.

\subsection{{\it no uncertainty} ($\mu\ra 0$) {\it un-regulated} 
($\lbd\ra\infty$)}

Equation (\ref{18}) becomes
\be
\label{29}
\frac{dy}{dt} = \al x + \bt y + By^3 \; ,
\ee
which is known as the Rayleigh equation. Notwithstanding the absence
of uncertainty on the fundamental value, this equation (\ref{29}) describes 
a market which allows for speculation on mispricing (for $\al>0$) and for 
nonlinear momentum trading (for $B>0$). This is a priori surprising as the 
idea that markets aggregate and disseminate information and also resolve 
conflicts is central to the literature on decentralization (Hurwicz, 1972) 
and rational expectations (Lucas, 1972). But, Plott and Sunder (1988) have 
shown that changing market institutions and trading instruments may lead
to deviations from rational equilibrium. Other reasons are the limits of 
arbitrage (Shleifer and Vishny, 1997): real arbitrage of mispricing require 
capital and entails risks, which makes mispricing difficult to eliminate
in certain circumstances. In our model, arbitrage is prevented when a 
majority of traders are speculative ($\al>0$) and/or follow nonlinear 
momentum strategies ($B>0$). The fact that a collective certainty on the 
fundamental value ($\mu\ra 0$) can coexist with speculative behavior and 
momentum trading is puzzling but should not surprise. For instance, it is 
well-documented that there exist situations in which people simply fail to 
do what is best even for themselves, in the face of good, freely available 
information. Despite stern warnings and mountains of strong evidence, some 
people continue to take up smoking (The McDonnell Social Norms Group, 2001), 
probably due to their less-than-perfect risk perception associated with the 
available information and their over-optimism about their longevity and 
future health (Sloan et al., 2003). Thus, information should be distinguished 
from the individual decision process which may be in contradiction with the 
direction that the information would favor. Information should also be 
distinguished from the aggregate of the individual decision processes which 
may be even less related to the public information due to possible feedbacks 
generated by interactions between people. A public information on the 
fundamental value of a stock may not be fully perceived as a reliable 
predictor of the market value because investors may have developed 
anticipation of the action of other investors which differ from the 
consequences of the public information. For instance, agent-based models 
of interacting investors who are subjected to their own private information,
to the public information and to the influence of their network of 
colleagues, may at times deviate from fundamental trading guided by the 
public information due to the influence of the other factors as well as 
their historical belief in the relative relevance of the different factors 
(Sornette and Zhou, 2006; Zhou and Sornette, 2006). This justifies to 
consider equation (\ref{15}) as a realistic limiting case representing a 
possible aggregate behavior of investors.

Only one fixed point $\{ 0,0\}$ exists for equation (\ref{29}). For $\al<0$ 
and $B<0$, the system is globally stable, with all trajectories converging 
to the stable point $\{ 0,0\}$ of an efficient market. When $\al<0$ but $B>0$, 
the point $\{ 0,0\}$ is stable, however its basin of attraction is finite, 
so that the system is globally unstable. The appearance of speculative 
traders $(\al>0)$ destroys the market for initial conditions or disturbances 
that bring the mispricing sufficiently far from zero, i.e., when the market 
price is sufficiently far from the fundamental price, for which mispricing 
trajectories become divergent.

For the sake of completeness, let us mention that, in the artificial case 
where we consider that the market friction coefficient $\bt$ can become 
positive, but $\al<0$ and $B<0$, then the point $\{ 0,0\}$ becomes unstable 
and there appears a limit cycle according to the Poincar\'e-Bendixon theorem 
(Hofbauer and Sigmund, 2002).

\section{Analysis of General Dynamics}

We now consider the general evolution equations (\ref{16},\ref{18}) 
with finite values of the parameters $\mu$ and $\lbd$ characterizing 
the market uncertainty and liberalization, respectively. As in the 
previous section, our analysis is performed in the limit where the 
stochastic component $\sigma dW$ is neglected, in order to explore 
the underlying nonlinear structure of the different dynamical regimes. 

First of all, we need to understand what are the conditions for the existence 
of stable fixed points and of other stationary regimes, if available. A first 
tool is linear analysis. Linear analysis gives information only on fixed 
points. In addition, there may exist limit cycles. However, as is well known, 
there is no general method for the determination of limit cycles. Though it 
is possible to formulate sufficient conditions for the existence of limit 
cycles, these conditions have the form of partial differential equations, 
which, anyway, require themselves a numerical solution (Giacomini and Viano, 
1995). It is then more straightforward to solve numerically the given system 
of ordinary differential equations, which we shall do to obtain an exhaustive 
description of the whole phase portraits. Below, we thus present the results
of the linear analysis, which includes both the determination of the fixed
points and their stability analysis. We complement the linear analysis
with a complete numerical study of the limit cycles, if any, and of their
stability.

\subsection{Linear analysis}

The fixed points of the dynamical system (\ref{16},\ref{18}) are given by the 
equations
\be
\label{30}
\al x + A x^3 \exp \left ( - \; \frac{x^2}{\mu^2} \right ) = 0 \; , 
\qquad y=0 \; ,
\ee
whose solutions will be denoted as $x^*$ and $y^*$. One evident solution is
\be
\label{31}
x_1^* = 0 \; , \qquad y_1^* = 0 \; ,
\ee
corresponding to an equilibrium efficient market with no mispricing, when 
the market price equals the fundamental value.

There can also exist two other fixed points $\{ x_2^*,0\}$ and 
$\{ x_3^*,0\}$, where $x_{2,3}^*$ are the solutions to the equation
\be
\label{32}
\al  + A x^2 \exp \left ( - \; \frac{x^2}{\mu^2} \right ) = 0 \; .
\ee
These additional solutions are possible only when the uncertainty of the 
fundamental value is large enough, i.e., for
\be
\label{33}
\mu \ge \mu_c \equiv \sqrt{-\; \frac{\al e}{A}} \; ,
\ee
and when the parameters $\al$ and $A$ have opposite signs:
\be
\label{34}
\frac{\al}{A} < 0 \; .
\ee
At the threshold value $\mu_c$,
\be
\label{35}
x_{2,3}^* = \pm \mu_c \qquad (\mu=\mu_c) \; .
\ee
With increasing uncertainty $\mu$, the absolute values of $x_{2,3}^*$ 
decrease. For large $\mu$, we have
\be
\label{36}
\left ( x_{2,3}^* \right )^2 \simeq -\; \frac{\al}{A} + 
\frac{\al}{\mu^2 A^2}  \qquad (\mu\gg 1) \; ,
\ee
so that
\be
\label{37}
\lim_{\mu\ra\infty} x_{2,3}^* = \pm \sqrt{-\; \frac{\al}{A}} \; .
\ee
Therefore, the inequality
\be
\label{38}
\left ( \frac{x_{2,3}^*}{\mu} \right )^2 \leq 1 \qquad 
( \mu_c \leq \mu < \infty )
\ee
is valid in the whole region $\mu\geq\mu_c$, where the fixed points 
$x_{2,3}^*$ exist.

The stability of the fixed points is determined by the properties of the 
Jacobian matrix $\hat J =[J_{ij}]$ which, in the present case, possesses 
the elements
$$
J_{11} \equiv \frac{\prt\dot{x}}{\prt x} = 0 \; , \qquad 
J_{12} \equiv \frac{\prt\dot{x}}{\prt y} = 1 \; ,
$$
\be
\label{39}
J_{21} \equiv \frac{\prt\dot{y}}{\prt x} = \al + 3A x^2 
\left ( 1 \; - \; \frac{2x^2}{3\mu^2} \right ) \exp\left ( - \; 
\frac{x^2}{\mu^2} \right ) \; ,
\ee
$$
J_{22} \equiv \frac{\prt\dot{y}}{\prt y} = \bt + 3B y^2 \left (
 1 \; - \; \frac{2y^2}{3\lbd^2} \right ) \exp\left ( - \; 
\frac{y^2}{\lbd^2} \right ) \; ,
$$
where the overdot corresponds to taking the time derivative. The 
eigenvalues of $\hat J$ are
\be
\label{40}
J^\pm = \frac{1}{2} \left [ {\rm Tr}\hat J \pm \sqrt{\left ( {\rm Tr}
\hat J \right )^2 - 4 {\rm det} \hat J} \right ] \; .
\ee
Since 
\be
\label{41}
{\rm Tr}\hat J = J_{22}\; , \qquad {\rm det} \hat J = - J_{21} \; ,
\ee
we get
\be
\label{42}
J^\pm = \frac{1}{2} \left ( J_{22} \pm 
\sqrt{J_{22}^2 + 4J_{21}} \right ) \; .
\ee
At the first fixed point $\{ x_1^*,y_1^*\} =\{ 0,0\}$, the eigenvalues 
$J^\pm$ read
\be
\label{43}
J^\pm_1 = \frac{1}{2} \left ( \bt \pm \sqrt{\bt^2 + 4\al} \right ) \; .
\ee
For the fixed points $\{ x_{2,3}^*,0\}$, we find
\be
\label{44}
J^\pm_{2,3} = \frac{1}{2} \left \{ \bt \pm \sqrt{\bt^2 - 
8 \left [ 1 - \left ( \frac{x_{2,3}^*}{\mu} \right )^2 \right ] } 
\right \} \; .
\ee

First, what can be immediately noticed is that, for the fixed points 
to be stable, the friction coefficient $\bt$ must be negative, as it has 
been assumed above. If the friction parameter was positive, there would 
be no stable fixed points. 

If $\al>0$ and $A>0$, the point $\{ 0,0\}$ is unstable, being a saddle 
point, and the fixed points $\{ x_{2,3}^*,0\}$ do not exist. This leaves 
us with three possible regimes for the signs of $\al$ and $A$, for which 
there can exist stable fixed points. To be more concise and precise, we 
classify these different cases by ascribing to each of them a name related 
to the signs of $\al$ and $A$. Recall that $\al<0$ corresponds to the 
mean-reversal behavior of individual traders, while $\al>0$, to their 
speculative behavior. One of the collective behaviors of traders, which 
is characterized by the parameter $A$, can also be either mean-reverting 
($A<0$), or speculative ($A>0$). This explains the names in the following 
classification
\begin{itemize}
\item 
$\al<0$, $A<0$: individual and collective mean-reverting market;
\item  
$\al>0$, $A<0$: individual speculative and collective mean-reverting market;
\item 
$\al<0$, $A>0$: individual mean-reverting and collective speculative market.
\end{itemize}
Below, we analyze in detail these three cases.

\subsection{Individual and collective mean-reverting market ($\al < 0$ and 
$ A < 0$)} 

Only one stable fixed point $\{ 0,0\}$ exists. It is a stable node if
$$
 -\; \frac{\bt^2}{4} \leq \al < 0 \; ,
$$
or a stable focus when
$$
\al < -\; \frac{\bt^2}{4}  \; .
$$
In the latter case, the approach to equilibrium is oscillatory with the 
effective asymptotic frequency
$$
\om_1 = \sqrt{4|\al|-\bt^2} \; .
$$

For finite $\mu$ and $\lbd$, the dynamics is much less trivial than 
in the limiting cases of the previous Section 3. There appears a number 
of rather unusual bifurcations. For given parameters $\al<0$ and $A<0$, 
distinct dynamics occur as a function of the amount of public information 
available on the fundamental price quantified by the parameter $\mu$. 
Specifically, a critical threshold $\mu_1$ separates two regimes, 
$\mu<\mu_1$ versus $\mu>\mu_1$, where the threshold value $\mu_1$ 
is a function of $\al$ and $A$.

For $\al < 0$ and $ A < 0$, all dynamical trajectories are bounded.
Although the unstable set of boundaries between different basins 
of attraction of the various fixed points and limit cycles can be quite 
nontrivial, especially for the uncertain markets, all trajectories tend 
to one of the attracting sets, either to the stable point $\{ 0,0\}$ or 
to a stable limit cycle. Since all trajectories are finite, the dynamical 
system is Lagrange stable. The stable set, including all basins of 
attraction, covers almost all phase plane, excluding the unstable set, 
having zero measure. In that sense, the dynamics is globally stable.

\subsubsection{Market with strong public information  
($0<\mu<\mu_1$)}

There exists a level of liberalization $\lbd_1$ such that, for 
$\lbd<\lbd_1$, there is only the stable fixed point $\{ 0,0\}$, as 
is shown in Fig. 1. Here and in what follows, the phase portrait is 
defined in the $x-y$ plane. But for more liberal markets ($\lbd>\lbd_1$), 
in addition to the fixed point $\{ 0,0\}$, which remains stable, there 
appears a stable limit cycle surrounding it, as shown in Fig. 2. This 
type of bifurcation is quite unusual. In the standard Hopf bifurcation, 
the limit cycle arises around an unstable fixed point (Hofbauer and 
Sigmund, 2002). But in our case, the point $\{ 0,0\}$ remains stable, 
while the new stable limit cycle appears. For $\lbd>\lbd_1$, the fixed 
point $\{ 0,0\}$ has a finite domain of attraction which is embedded 
within the domain of attraction of the limit cycle. The appearance of 
the latter is due to the existence of trend followers for $B>0$. If the 
sign of $B$ changes to negative, the limit cycle disappears. The further 
increase of the liberalization parameter $\lbd$ leaves the topological 
picture unchanged, with one stable point $\{ 0,0\}$ and one stable limit 
cycle around it. The basin of attraction of the point $\{ 0,0\}$ does not 
change much, but the size of the limit cycle rapidly grows with increasing 
$\lbd$. This is illustrated in Fig. 3, which can be compared with Fig. 2.

\subsubsection{Uncertain market ($\mu_1 < \mu < \infty$)}

In this case, there exist three liberalization thresholds 
$\lbd_1<\lbd_2<\lbd_3$. For strictly regulated market, $\lbd<\lbd_1$, 
there is just one stable point $\{ 0,0\}$, as in Fig. 4. When the degree 
of liberalization lies in the interval $\lbd_1<\lbd<\lbd_2$, one stable 
point $\{ 0,0\}$ is coexisting with a limit cycle surrounding it, as in 
Fig.5. In the interval $\lbd_2<\lbd<\lbd_3$, the unique limit cycle is 
replaced by two stable limit cycles, one internal and the other external, 
both surrounding the point $\{ 0,0\}$, as shown in Fig. 6. Increasing 
further the liberalization $\lbd>\lbd_3$ breaks the internal limit cycle, 
but the external cycle remains and grows, as shown in Fig. 7. The limit 
cycles are due to the existence of nonlinear trend followers: for $B<0$, 
all limit cycles disappear.

\subsection{Individual speculative ($\al>0$) and collective 
mean-reverting ($A<0$) market}

Let us consider the case where individual traders are speculative
$(\al>0)$, but collectively mean-reversal $(A<0)$. If there is little
uncertainty on the fundamental price ($\mu<\mu_c$, with $\mu_c$ defined
in Eq.~(\ref{33}), then there are no finite attractors of the dynamics, 
and all trajectories diverge. This actually implies that the market does 
not exist. Only when $\mu>\mu_c$, two stable fixed points $\{ x_{2,3}^*,0\}$ 
exist, which are the solutions of Eq. (\ref{32}). These fixed points are 
stable nodes when $\bt^2 \geq \bt_{2,3}^2$ and stable focuses when 
$\bt^2 < \bt_{2,3}^2$, where 
\be
\label{45}
\bt_{2,3} \equiv 8\al \left [ 1 - \left ( \frac{x_{2,3}^*}{\mu} 
\right )^2 
\right ] \;.
\ee

There is another threshold value $\mu_1>\mu_c$ separating different 
dynamics. Also, there exists a speculative threshold $\al_c>0$, below and 
above which dynamics is different. The most diverse possibilities occur for 
large uncertainty $(\mu>\mu_1)$ and strongly speculative $(\al>\al_c)$ 
markets, which present four  regimes, depending on the level of regulation 
quantified by the liberalization parameter $\lbd$. Less speculative 
$(\al<\al_c)$ and less uncertain $(\mu<\mu_1)$ markets exhibit only a part 
of these regimes. It is therefore convenient to start the investigation 
from the most ramified case.

As shown in the classification presented in the next subsections, the
general feature of this class of markets with individual speculation and
collective mean-reverting properties is that the dynamics is only locally
stable. There always exist the basins of infinity preventing a global market 
stability.

\subsubsection{Uncertain market $(\mu_1<\mu<\infty)$ with strong speculation
$(\al_c<\al<\infty)$}

There are three liberalization thresholds, $\lbd_1<\lbd_2<\lbd_3$ and four 
dynamic regimes. When the market is strictly regulated $(0<\lbd<\lbd_1)$, 
there are two stable fixed points $\{ \pm s,0\}$. Here we again use the 
short-hand notation $x_{2,3}^*\equiv\pm s$, with $s>0$. The related phase 
portrait is shown in Fig. 8. The point $\{ s,0\}$ corresponds to a positive 
convention, while $\{-s,0\}$ corresponds to a negative convention. The 
basins of attraction of these points have infinite measure, however they 
do not cover the whole phase plane. To their left and right lie the basins 
of infinity, where all trajectories diverge. Increasing the market 
uncertainty $\mu$ increases the basins of attraction of $\{ \pm s,0\}$, 
but the basin of infinity disappears only in the limit of completely 
uncertain markets, $\mu\ra\infty$, which reduces to the Duffing equation 
studied in Section \ref{ngnlflvq}. Thus, the basin of infinity, with 
infinite measure, exists for any finitely uncertain market, no matter how 
large $\mu$ is.

For weaker regulations, in the interval of liberalization 
$\lbd_1<\lbd<\lbd_2$, there are two speculative stable points $\{ \pm s,0\}$ 
and two stable limit cycles around each of them, as shown in Fig. 9. There 
are four basins of attraction, but their union again does not cover the 
whole plane, since there are two basins of infinity.

For still more unregulated markets, in the interval of liberalization 
$\lbd_2<\lbd<\lbd_3$,  two separate limit cycles of the previous regime 
merge into one big stable cycle surrounding both fixed points, as shown 
in Fig. 10. Again, their total basins of attraction do not cover the 
whole phase plane. When the market is weakly regulated, such that 
$\lbd_3<\lbd<\infty$, the limit cycle of the previous regime breaks up, 
and only the two stable fixed points $\{ \pm s,0\}$ remain. The resulting 
phase portrait is depicted in Fig. 11. The basin of attraction of these 
fixed points does not cover the whole phase plane.

\subsubsection{Uncertain market $(\mu_1<\mu<\infty)$ with weak 
speculation $(0<\al<\al_c)$}

In this case, there exist two liberalization thresholds, $\lbd_1<\lbd_2$, 
and three dynamic regimes. When the market regulation is strict, 
$0<\lbd<\lbd_1$, there are two stable points, $\{ \pm s,0\}$, similar to 
the situation shown in Fig. 8. When the liberalization parameter lies in 
the interval $\lbd_1<\lbd<\lbd_2$, there appear a stable limit cycle around 
these points, analogously to the case shown in Fig. 10. And for a strongly 
deregulated market, with $\lbd_2<\lbd<\infty$, there exist just two stable 
points $\{ \pm s,0\}$, and no cycles, as in the case shown in Fig. 11. In 
summary, the weakly speculative $(\al<\al_c)$ uncertain market differs from 
the strongly speculative $(\al>\al_c)$ one by the absence of the intermediate 
regime with two points and two cycles, which occurs for the latter market, as 
seen in Fig. 9. The phase plane again consists of the union of the basins of 
attraction and of two basins of infinity.

\subsubsection{Intermediate uncertainty market $(\mu_c<\mu<\mu_1)$}

The phase portrait is topologically the same for any $\al>0$ and $\lbd$.
There exist only two stable fixed points $\{ \pm s,0\}$. No cycles
appear. The dynamics is analogous to that illustrated in Fig. 8. The
basin of attraction never covers the whole phase plane. Recall that the
fixed points exist only if $\mu>\mu_c$. A speculative $(\al>0)$ market,
with rather strong certainty $\mu<\mu_c$, cannot exist at all, since all
trajectories diverge.

\subsection{Individual mean-reverting $(\al<0)$ and collective 
speculative $(A>0)$ market }

Such a situation can develop, when the majority of traders on the 
average are rational, but there are groups of speculators whose action 
can eventually dominate at times of large deviations from the fundamental 
value. Again, the level of public uncertainty in the market concerning 
the fundamental price drives predominantly its dynamics. There are two 
thresholds, $\mu_c$ and $\mu_1$, characterizing the market uncertainty. For 
each given $\mu$, the dynamics depend on the level of liberalization $\lbd$.

All the various regimes of collectively speculative markets with individual
mean-reverting behavior are globally stable. All trajectories go either to 
one of the stable fixed points or to one of the stable limit cycles, as we 
now classify.

\subsubsection{Small uncertainty on the fundamental value 
$(0<\mu<\mu_c)$}

When the market is strictly regulated, such that $0<\lbd<\lbd_1$, there 
exists just one stable point $\{ 0,0\}$. The phase portrait is shown in 
Fig. 12. The basin of attraction is the whole phase plane.

If the well-informed market is weakly regulated, so that 
$\lbd_1<\lbd<\infty$, there appears a stable limit cycle around the 
stable point $\{ 0,0\}$ as illustrated by Fig. 13. There are two basins 
of attraction, one corresponding to the stable point and the other one to 
the cycle. The total stable set, consisting of the union of these basins 
of attraction, covers the whole phase plane. The unstable set, composed of 
the boundaries between the basins of attraction, has zero measure.

Increasing the liberalization parameter $\lbd$ leaves the topology of the 
phase portrait similar to that shown in Fig. 13, with one stable point 
$\{ 0,0\}$ and a stable cycle around it. However, a slight increase of the 
liberalization $\lbd$ substantially increases the limit cycle and deforms 
its shape, as shown in Fig. 14.

\subsubsection{Intermediate level of uncertainty $(\mu_c<\mu<\mu_1)$}

As soon as the uncertainty $\mu$ on the fundamental value overpasses 
the threshold $\mu_c$ defined in Eq.~(\ref{19}), there appear three stable 
fixed points, $\{ 0,0\}$ and $\{ \pm s,0\}$. For strictly regulated markets 
($0<\lbd<\lbd_1$), only these three points exist for a market characterized 
by an intermediate uncertainty $(\mu<\mu_1)$. The corresponding phase 
portrait is shown in Fig. 15. The total stable set covers the whole phase 
plane.

For the market with intermediate uncertainty and weak regulation
($\lbd_1<\lbd<\infty$), in addition to the same three stable points just 
mentioned, a stable limit cycle appears which surrounds all of them, as 
represented in Fig. 16. The total stable set, as before, covers the whole 
phase plane.

\subsubsection{Uncertain market $(\mu_1<\mu<\infty)$}

There are two thresholds of market liberalization, $\lbd_1<\lbd_2$. For
a strictly regulated market ($0<\lbd<\lbd_1$), there are three stable
fixed points, $\{ 0,0\}$ and $\{ \pm s,0\}$. The phase portrait is
analogous to that shown in Fig. 15, with the only difference that the
points $\{ \pm s,0\}$ are shifted away from the center $\{ 0,0\}$.

For intermediate regulation ($\lbd_1<\lbd<\lbd_2$), the phase portrait 
changes, with the existence of three fixed points, $\{ 0,0\}$ and 
$\{ \pm s,0\}$ and two stable limit cycles around each of the latter 
points $\{ \pm s,0\}$. This is shown in Fig. 17. The whole phase plane 
is covered by the total stable set.

With the increase of the liberalization parameter $\lbd$ in the interval 
$\lbd_2<\lbd<\infty$, the two limit cycles of the previous regime break 
up and combine into one big limit cycle surrounding all three fixed 
points. The corresponding phase portrait is shown in Fig. 18. As in the 
previous cases, except for the boundaries between the different domains 
of attraction which form a set of zero measure, the total stable set 
covers the whole phase plane.

\subsection{Summary of the classification of different market 
types}

The classification presented in the previous sections demonstrate a 
very large variety of possible markets. There are 6 kinds of markets 
with individual and collective mean-reverting behaviors, 8 variants 
of markets with individual speculative and collective mean-reverting 
behaviors, and 7 types of markets with individual mean-reverting and 
collective speculative behaviors. Thus, in total, there exist 21 types 
of markets. In order to better compare these different markets, we 
summarize their properties in Tables 2, 3, and 4. As explained above, 
this classification sorts out markets according to three basic types:
\begin{itemize}
\item 
the individual and collective mean-reverting markets $(\al<0,\; A<0)$ 
(Table 2);
\item 
the individual speculative and collective mean-reverting markets 
$(\al>0,\; A<0)$ (Table 3);
\item 
the individual mean-reverting and collective speculative markets
$(\al<0,\;  A>0)$ (Table 4). 
\end{itemize}
Then, a second level of classification is performed on the basis 
of the level of liberalization. For brevity, we refer to a strictly 
regulated market as {\it rigid} and to a weakly regulated market as 
{\it soft}. Intermediate cases are termed mid-rigid or mid-soft with 
a meaning which is clear from the context. In the right columns of the 
Tables, the corresponding attractors are shown. The stable fixed points 
are denoted as $\{ 0,0\}$ and $\{ \pm s,0\}$. The notation $C_0$ implies 
a stable limit cycle surrounding the point $\{ 0,0\}$ and $C_0'$ is 
another limit cycle around this point. The stable cycles $C_s$ and 
$C_{-s}$ are around the points $\{ s,0\}$ and $\{ -s,0\}$, respectively. 
The limit cycle $C_{\pm s}$ surrounds both these points $\{ \pm s,0\}$. 
Finally, the notation $C_{0,\pm s}$ represents a stable limit cycle 
surrounding all three points, $\{ 0,0\}$ and $\{ \pm s,0\}$.

With respect to their topological properties, not all markets are 
distinct in our classification presented in the Tables 2-4. In other 
words, not all markets presented as different in the Tables 2-4 possess 
topologically different phase portraits. However, the classification 
presented in Tables 2-4 is convenient for practical purposes, as it 
distinguishes both quantitatively and qualitatively different types 
of markets with respect to their characteristics in terms of level of 
uncertainty and of regulation. Markets with low uncertainty and strict 
regulation display the simplest dynamics. The complexity of the dynamics 
increases for more uncertain and less regulated markets. The appearance 
of speculative traders in uncertain markets makes their dynamics quite 
nontrivial, even if the regulation is rather strict. In this way, the 
speculative points $\{ \pm s,0\}$ appear, corresponding to positive or 
negative conventions. The existence of trend followers leads to the 
occurrence of limit cycles.

\section{Study of Stochastic Dynamics}

The full dynamics described by equations (\ref{16},\ref{18}) involves 
in addition the stochastic term $\sigma dW$. As explained in section 
2.4, the interplay of the structure of attractors of the nonlinear 
deterministic dynamics with the stochasticity $\sigma dW$ can be
expected to provide a dynamic underpinning for many stylized facts 
and properties of financial prices. As we have discussed, the nonlinear
deterministic dynamics embodies the interplay between the aggregate
effects of several market and agent contributing factors. Complementarily, 
stochasticity models both the granularity decorating these different 
contributions as well as exogenous perturbations such as news and 
sunspots. This section provides a first numerical exploration of 
some of the more interesting properties for which we offer a qualitative 
intuitive interpretation based on the classification of the 21 market 
regimes obtained above.

To investigate the influence of stochasticity, we solve the stochastic 
differential equations (\ref{16}) and (\ref{18}) numerically by employing 
the Euler discretization scheme. Then the discretized version of 
Eq.~(\ref{16}) reads
\be
x(t_i + \Dlt t) - x(t_i) = y(t_i) \Dlt t + \sgm\; 
\sqrt{\Dlt t}\;  R(t_i)
\ee
where $t_i$, with $i=1,2,\ldots$ are the discrete times in units of 
the time step $\Dlt t$. The random numbers $R(t_i)$ are generated with 
the standard Gaussian distribution (zero mean and unit variance). The 
amplitude of stochasticity is controlled by the volatility parameter 
$\sgm$. Below, we present several figures illustrating the influence 
of increasing $\sgm$ on the mispricing trajectories $x(t)$. Recall that, 
by definition (\ref{3}), $x(t)$ is the log-price $\log p(t)$ shifted by 
the fundamental log-value. We examine in turns some of the most 
illustrative cases.

Strictly regulated markets with low uncertainty exhibit a rather 
simple dynamics and, as such, their qualitative properties are the 
least modified by the presence of stochasticity. For such markets 
with mean-reversing investors $(\al<0, A<0)$ whose phase portrait 
is shown in Fig. 1, the trajectory $x(t)$, in the presence of 
stochasticity, follows an approximate Ornstein-Uhlenbeck process 
(random walk with mean reversal) around the efficient market state 
$\{ 0,0\}$, slightly modified by the presence of the nonlinear terms. 
While this may provide a source for some on the nonlinearities that 
have been documented (Hsieh, 1995), much more interesting and 
complicated price trajectories appear for markets characterized by the 
coexistence of several attractors.

Let  us consider, for instance, a weakly regulated market with low 
uncertainty and all mean-reverting agents $(\al<0, A<0)$. Its phase 
portrait, given in Fig. 3, shows the co-existence of the equilibrium 
fixed point $\{ 0,0\}$ with a non-linear periodic cycle. Typical 
trajectories, starting with the same initial conditions $x(0)=0$, 
$y(0)=1$, under increasing volatility $\sgm$, are presented in Fig. 19. 
In the absence of volatility, the initial momentum $y(0)=1$ is small 
enough that the intial conditions belong to the domain of attraction 
of the equilibrium fixed point $\{ 0,0\}$, so that the mispricing $x(t)$ 
converges to the efficient market state, as shown in Fig. 19 (a). In the 
presence of a small volatility $\sgm=0.1$, typical trajectories converge 
to a neighborhood of $\{ 0,0\}$, and then exhibit an approximate 
Ornstein-Uhlenbeck process around the efficient market state $\{ 0,0\}$, 
slightly modified by the presence of the nonlinear terms, as shown in 
Fig. 19 (b). A typical trajectory for a larger volatility $\sgm=0.5$ 
leads to shifts between the domain of attraction of the fixed point 
$\{ 0,0\}$ to the nonlinear cycle. In the example shown in Fig. 19 (c), 
up to $t=10$, the mispricing is undergoing an approximate 
Ornstein-Uhlenbeck process around the efficient market state $\{ 0,0\}$. 
After this time, one can observe a transition to a quasi-periodic cycle 
associated with the domain of attraction of the large nonlinear cycle. 
Since the domain of attraction of the later is much larger than that of 
the fixed point  $\{ 0,0\}$, a return to approximate Ornstein-Uhlenbeck 
process around the efficient market state $\{ 0,0\}$ is not excluded but 
will be rare for the choice of parameters of this figure. For even larger 
volatility $\sgm=5$ as shown in Fig 19 (d), the mispricing is following 
a quasi-periodic random walk, with intermittent trapping close to the 
efficient market state $\{ 0,0\}$ followed by re-captures by the 
quasi-periodic random walk. Such mispricing could be a good representative 
of noisy business cycles (Tvede, 2006).

For markets characterized by more than just two basins of attraction,
the presence of stochasticity brings in the possibility for some domains 
of attractions to be destroyed by fusion with others, while different 
domains of attractions may survive and keep their qualitative shape and 
properties. Conside for instance the case of slightly deregulated markets 
with large uncertainty, with individual speculative and collective 
mean-reverting agents, whose phase portrait in the absence of 
stochasticity is shown in Fig. 9. By showing typical trajectories, 
Fig. 20 illustrates the influence of increasing the volatility. In 
the absence of stochasticity, the phase portrait of Fig. 9 shows the 
existence of two stable fixed points and two stable limit cycles around 
each of the points. In the absence of stochasticity $\sgm=0$, any 
trajectory, starting inside the basin of attraction of one of the fixed 
points, converges to the latter as shown in Fig. 20 (a). For weak 
stochasticity ($\sgm=0.1$), typical mispricing trajectories exhibit 
rather regular oscillations, suggesting a kind of resonance, as shown 
in Fig. 20 (b). With a slightly stronger volatility ($\sgm=0.3$), typical 
trajectories first oscillate around the fixed point before passing over 
to the domain of attraction of the limit cycle surrounding the fixed 
point, as shown in Fig. 20 (c). In that sense, the basin of attraction 
of the fixed point has been destroyed, though the limit cycle has not 
been changed much, the cycles being decorated by a random noise, forming 
a kind of stochastic trajectory co-integrated with the initial unperturbed 
cycle. With large volatility ($\sgm=3$) as shown in Fig. 20 (d), one can 
observe jumps between the two noisy cycles, one (resp., the other) that 
can be interpreted as a positive (resp. negative) business cycle with a 
valuation above (resp. below) the fundamental price.

The gradual destruction of the basins of attraction, under increasing
volatility, is a general phenomenon existing for different kinds of
attractors. This is demonstrated in Figs. 21 and 22 for two different
markets characterized by the phase portraits of Fig. 15 and Fig. 17,
respectively, in the absence of stochasticity. The former case
corresponds to a strictly regulated market with intermediate
uncertainty, exhibiting three stable fixed points. With increasing
volatility, one can observe that typical trajectories that start in the
basin of attraction of one of the fixed points, as in Fig. 21 (a), are
mildly perturbed with a mean-reversion stochastic behavior around 
the fixed point over long time periods for weak volatility ($\sgm=0.2$)
as in Fig. 21 (b), but can jump eventually to the basin of attraction of 
the neighboring fixed points. The random waiting times between jumps 
become shorter for larger volatility as illustrated in Fig. 21 (c), 
which shows a jump from a stochastic dynamics in the neighborhood of 
the positive convention fixed point to a stochastic dynamics in the 
vicinity of the equilibrium fixed point. At longer time scales and/or 
for larger volatility, one can observe alternate jumps between the three 
fixed points with transient random excursions around each of these fixed 
points, as shown in Fig. 21 (d). Technically, the mathematical theory 
describing the statistics of the waiting times between such jumps can 
be developed, based on the concept of ``effective action'' (Eyink, 1998), 
which generalizes reaction rate theory or the Kramers' problem (H\"anggi 
et al., 1990) for the case where the dynamics is Hamiltonian to the 
non-variational case. This behavior involving random jumps between 
distinct stochastic phases associated with different underlying fixed 
points or attractors is reminiscent of the regime shifts between 
equilibrium to positive or negative conventions and between them.

The rich phase structure of markets with intermediate regulation and
large uncertainty, whose phase portrait is given in Fig. 17, evolves 
with increasing volatility by a progressive destruction of the attractors 
which occur in several stages. There are initially three fixed points and 
two limit cycles around two of the points. A typical trajectory, starting 
close to one of the fixed points inside the related cycle, as in Fig. 22 
(a) ($\sgm=0$), oscillates around this point for intermediate values of 
the volatility ($\sgm=1$), as shown in Fig. 22 (b). For larger volatility 
($\sgm=3$), the fixed point loses its attraction and the stochastic 
trajectory evolves to a random noise decorating the limit cycle
surrounding it, as in Fig. 22 (c). With the increase of the volatility 
to large values ($\sgm=5$), typical mispricing trajectories alternate 
randomly between the two former limit cycles, with random added structures, 
as in Fig. 22 (d). The stochastic nonlinear dynamics described in Figs. 21 
and 22 provides a natural mechanism for regime shifts observed in real 
financial markets. 

It is worth stressing that the basins of attraction of limit cycles, 
when they appear, are practically always larger than the basins of 
attraction of the coexisting fixed points. As a consequence, increasing 
the volatility destroys first the basins of the fixed points and only 
later the basins of the limit cycles. In that sense, the limit cycles 
are more stable to increasing volatility, suggesting a mechanism by 
which business cycles play a dominant role in explaining the financial
price dynamics, while the relative explanatory power of fundamental 
pricing, and even of systematic stable deviations such as convention 
pricing, remains smaller.

A clear example of the situation in which increasing volatility 
destroys the basin of attraction of all fixed points while only 
disturbing a limit cycle is presented in Fig. 23. This corresponds
to a weakly regulated uncertain market characterized by the phase 
portrait shown in Fig. 18. With zero volatility, there are three fixed 
points and a single limit cycle surrounding them. Going from vanishing 
to non-zero volatility very quickly destroys the basins of attraction of 
all three fixed points. But stochastic oscillations in the neighborhood 
of the limit cycle persist for rather large volatilities, as illustrated 
in Fig 23 (b).

Long-time simulations exemplify the transitions between the trapping 
regions, which are the remnant structures associated with the fixed 
points and limit cycles in absence of stochasticity. Figs. 24 and 25 
show two typical trajectories of the mispricing for two different 
markets, over a timespan an order of magnitude larger than in the 
previous Figs. 19 to 23. Fig. 24 shows a typical trajectory of the 
mispricing $x(t)$ for a strictly regulated market with intermediate 
uncertainty, corresponding to the phase portrait of Fig. 15 in absence 
of stochasticity, for a volatility $\sgm=2$. The stochasticity induces 
random transitions between mostly the three fixed points, equilibrium 
($\{ 0,0\}$) and the positive and negative conventions ($\{-s,0\}$ and 
$\{ s,0\}$). The maximal number of attractors, which is five in our 
model, is obtained for uncertain weakly regulated markets with 
individual reverting and collectively speculating agents, corresponding 
to the phase portrait shown in Fig. 17. Under the influence of 
stochasticity, the typical trajectory depicted in Fig. 25 jumps between 
the two limit cycles around the convention fixed points $\{-s,0\}$ and 
$\{s,0\}$. One can also observe that, inside each limit cycle phase, the 
trajectory also transiently oscillates around the corresponding fixed 
point. Oscillations around the point $\{ 0,0\}$ do not last long, because 
the trapping region around this point is smaller than the trapping regions 
of the limit cycles. Thus, the existence of even weak noise destroys the 
attractors, transforming the related basins of attraction into global 
trapping regions in which the dynamics is erratic. The local properties, 
at each given moment of time $t$, can be characterized by such 
time-dependent quantities as local stability indices and local expansion 
exponents (Yukalov, 2002, 2003). When the noise is very weak, a trajectory 
can practically forever live inside one of the trapping regions. But 
increasing noise induces the trajectories to jump more frequently between 
different trapping regions. The larger a trapping region, the longer is 
the time spent by the trajectory inside it. The transitions between 
different trapping regions correspond to transitions between different 
market regimes, such as the efficient-market state, positive or negative 
conventions, and business cycles. More detailed applications to particular 
markets will be studied in our subsquent papers.

In summary, stochasticity modifies the classification in terms of 
attractors consisting of fixed points and limit cycles, as follows. 
First, strictly speaking, attractors disappear and are replaced by
trapping regions, because stochasticity enables the trajectory to
eventually escape from any domain of attractions. However, it remains 
sensible to think of attractors in terms of trapping domains in which 
the stochastic trajectory remains confined for some significant time 
before eventually escaping, perhaps being reinjecting at a later time, 
and so on. The trapping time is all the longer, the smaller is the 
volatility. Thus, for weak stochasticity, the trapping domains are 
closely approximated by the basins of attractions and the stochastic 
dynamics can be approximated by random walks trapped around the 
attractors (fixed points and limit cycles) sometimes jumping to other 
attractors and so on. As the amplitude of the stochasticity is
increased, the partition of trajectories in terms of trapping and jumps 
become blurred for the fixed points, and especially the equilibrium fixed 
point which has often the smallest domain of attraction. In contrast, the 
trapping regions inherited from the basin of attractions of the limit 
cycles, when they exist, remain better defined in the sense that typical 
trajectories are found to remain trapped for significant times in such 
regions before jumps in fast time to other regions. For still stronger 
volatility, the trapping regions associated with the cycles eventually 
fuse together and typical trajectories become random walk-like with 
nonlinear chaotic modulations.

\section{Conclusion}

We have proposed novel evolution equations of nonequilibrium markets,
based on general symmetry and structural considerations. The obtained 
equations (\ref{16}, \ref{18}) take into account the rational or 
speculative behavior of individual traders, market friction, the 
possibility of collective speculative behavior, the existence of trend 
followers, the influence of uncertainty on the fundamental value and 
the level of market regulation. The dynamics resulting from these 
equations turns out to be extremely rich, yielding rather nontrivial 
phase portraits and a variety of different regimes, evolving from one 
to another through complex bifurcations. We have found 21 different 
types of markets. One of these markets is the efficient market 
equilibrium. But there are 20 other types of markets, some being 
characterized by positive or negative conventions, transient bubbles 
and crashes, as well as rather complex business cycles, which coexist 
with the equilibrium fixed point.

In general, for low and intermediate uncertainty, only fixed points
exist, with the equilibrium point possibly coexisting with a positive
and a negative convention. The corresponding stochastic dynamics is
co-integrated with the price corresponding to one of these fixed points,
but not for ever, as the coexistence between the three fixed points
implies that the dynamics jumps spontaneously from one regime to another
at random times. For large uncertainty and for weak regulations, limit
business cycles appear in addition, competing with the other fixed
points. The theory predicts transient co-integration of the observed
price with either the fundamental price, one convention price or a
regular nonlinear business cycle, followed by sudden jumps to another
transient co-integration with one of these regimes. As a consequence,
stochasticity blurs the basins of attractions, changing them into
transient trapping regions, and making these trapping regions
progressively merge when the intensity (volatility) of stochasticity
increases. 

Applications and calibration of this general theory to
particular markets will be developed in separate publications.

\vskip 5mm

{\bf Acknowledgement}

\vskip 2mm

We are very grateful to Y. Malevergne for helpful discussions and 
useful remarks.

\newpage

{\parindent=0pt

{\Large{\bf References}}

\vskip 5mm

Andersen, J.V., Gluzman, S., Sornette, D., 2000.
Fundamental framework for technical analysis. 
European Physical Jornal B 14, 579-601.

\vskip 2mm

Andersen, J.V., Sornette, D., 2004. Fearless versus fearful 
speculative financial bubbles. Physica A 337, 565-585.

\vskip 2mm

Arnold, V.I, 1978. Ordinary Differential Equations. Cambridge:
MIT Press.

\vskip 2mm

Bachelier, L., 1900. Th\'eorie de la Speculation. 
Paris: Gauthier-Villars. 

\vskip 2mm 

Black, F., 1976. Studies of stock price volatility changes. 
Proceedings of the American Statistical Association. Business and 
Economics Statistics Section, 117-181.

\vskip 2mm 

Bonomo, M., Garcia, R., 1994a. Can a well-fitted equilibrium 
asset pricing model produce mean reversion? Journal of Applied 
Econometrics 9, 19-29.

\vskip 2mm 

Bonomo, M., Garcia, R., 1994b. Disappointment aversion as a 
solution to the equity premium and the risk-free rate puzzles. 
Working Paper 94s-14, CIRANO.

\vskip 2mm 

Bonomo, M., Garcia, R., 1996. Consumption and equilibrium asset 
pricing: an empirical assessment. 
Journal of Empirical Finance 3, 239-265.

\vskip 2mm 

Bouchaud, J.P., Cont, R., 1998. Langevin approach to stock market 
fluctuations and crashes. European Physical Journal B 6, 543-550.

\vskip 2mm

Boyer, R.,  Orl\'ean, A., 1992. How do conventions evolve?
Journal of Evolutionary Economics 2, 165-177.

\vskip 2mm

Broekstra, G., Sornette, D., Zhou, W.X., 2005. Bubble, critical 
zone and the crash of Royal Ahold. Physica A 346, 529-560. 

\vskip 2mm

Cecchetti, S. G., Lam, P.,  Mark, N. C., 1990. Mean reversion in 
equilibrium asset prices.  American Economic Review 80, 398-418.

\vskip 2mm 

Cecchetti S. G., Lam, P., Mark, N. C., 1993. The equity premium 
and the risk free rate: matching the moments. Journal of Monetary 
Economics 31, 21-45.

\vskip 2mm 

Eymard-Duvernay, F., Favereau, O., Orl\'ean, A., Salais, R., 
Th\'evenot, L.,  2005. Pluralist  integration in the economic
and social sciences: the economy of conventions. Post-Autistic 
Economics Review 34, 22-40.

\vskip 2mm

Farmer, J.D., 2002. Market force, ecology and evolution. Industrial 
Corporate Change 11, 895-953.

\vskip 2mm

Figlewski, S., Wang, X., 2000. Is the "leverage effect" a leverage 
effect? Working Paper. New York: New York University.

\vskip 2mm

Giacomini, H., Viano, M., 1995. Determination of limit cycles for 
two-dimensional dynamical systems. Physical Review E 52, 222-228.

\vskip 2mm

Gluzman, S., Yukalov, V.I., 1997. Algebraic self-similar 
renormalization in the theory of critical phenomena. 
Physical Review E 55, 3983-3999.

\vskip 2mm

Gluzman, S., Yukalov, V.I., 1998. Booms and crashes in self-similar 
markets. Modern Physics Letters B 12, 575-587.

\vskip 2mm

Grassia, P.S., 2000. Delay, feedback and quenching in financial 
markets. European Physical Journal B 17, 347-362.

\vskip 2mm

Hamilton, J.D., 1989. A new approach to the economic analysis of
nonstationary time series and the business cycle. Econometrica
57, 357-384.

\vskip 2mm

Hamilton, J.D., Raj, B., 2002. New directions in business cycle 
research and financial analysis. Empirical Economics 27, 149-162.

\vskip 2mm

He, H., Leland, H., 1993. On equilibrium asset price processes. 
Review of Financial Studies 6, 593-617.

\vskip 2mm

Hofbauer, J., Sigmund, K., 2002. Evolutionary Games and 
Population Dynamics. Cambridge: Cambridge University. 

\vskip 2mm

Horsthemke, W., Lefever, R., 1984.
Noise-Induced Transitions: Theory and Applications in Physics, 
Chemistry and Biology. Berlin: Springer.

\vskip 2mm

Hsieh, D.A., 1995. Nonlinear dynamics in financial markets: 
evidence and implications.  Financial Analysts Journal 51, 55-62.
  
\vskip 2mm 

Hurwicz, L., 1972. On informationally decentralized systems. In:
McGuire, C.B., Radner, R. (Eds.), Decision and Organization.  
Amsterdam: North Holland, 297-336.

\vskip 2mm

Ide, K., Sornette, D., 2002. Oscillatory finite-time 
singularities in finance, population and rupture. Physica A 307, 
63-106.

\vskip 2mm

Lucas, R.E., 1972. Expectations and the neutrality of money. 
Journal of Economic Theory 4, 103-124.
    
\vskip 2mm

Orl\'ean A., 1994. Analyse Economique des Conventions. Paris: PUF.

\vskip 2mm

Pandey, R.B., Stauffer, D., 2000. Search for log-periodicity 
oscillations in stock market simulations. International Journal of
Theoretical and Applied Finance 3, 479-482.

\vskip 2mm

Plott, C.R., Sunder, S., 1988. Rational expectations and the 
aggregation of diverse information in laboratory security markets. 
Econometrica 56, 1085-1118.

\vskip 2mm

Shefrin, H., 2000. Beyond Greed and Fear: Understanding Behavioral 
Finance and the Psychology of Investing. 
Boston: Harvard Business School.

\vskip 2mm

Shiller, R.J., 1989. Market Volatility. Cambridge: MIT.

\vskip 2mm

Shleifer, A.,  Vishny, R.W., 1997. The limits of arbitrage.
The Journal of Finance 52, 35-55.

\vskip 2mm
	 
Sloan, F.A., Smith, V.K., Taylor, D.H., Jr., 2003.
The Smoking Puzzle Information, Risk Perception, and Choice.
Boston: Harvard University.

\vskip 2mm

Sornette, D., 1998. Discrete scale invariance and complex dimensions. 
Physics Reports 297, 239-270. 

\vskip 2mm

Sornette, D., 2000. Stock market speculation: spontaneous symmetry 
breaking of economic valuation. Physica A 284, 355-375. 

\vskip 2mm

Sornette, D., 2003. Critical market crashes.
Physics Reports 378, 1-98. 

\vskip 2mm

Sornette, D., 2003. Why Stock Markets Crash. 
Princeton: Princeton University.

\vskip 2mm

Sornette, D., 2004. Critical Phenomena in Natural Sciences. 
Berlin: Springer.

\vskip 2mm

Sornette, D., Andersen, J.V., 2002. A nonlinear 
super-exponential rational model of speculative financial bubbles. 
International Journal of Modern Physics C 13, 171-187.

\vskip 2mm

Sornette, D., Johansen, A., 2001. Significance of log-periodic 
precursors to financial crashes. Quantitative Finance 1, 452-471. 

\vskip 2mm

Sornette, D.,  Zhou, W.-X., 2006. Importance of positive 
feedbacks and over-confidence in a self-fulfilling Ising model 
of financial markets. Physica A 370, 704-726.

\vskip 2mm

The McDonnell Social Norms Group, 2001. Battling bad behavior. 
The Scientist 2.
	
\vskip 2mm

Thurner, S., 2001. Financial asset price dynamics: a dynamical 
thermostat model. 
Proceedings of American Institute of Physics 574, 60-69.

\vskip 2mm

Thurner, S., Dockner,  E.J., Gaunersdorfer, A., 2002. 
Asset price dynamics in a model of investors operating on 
different time horizons. SFB-Working Paper No. 93, 
Universit\"at Wien, e-print cond-mat/0011286.

\vskip 2mm

Tvede, L., 2006. Business Cycles. New York: John Wiley.

\vskip 2mm

von Neumann, J., Morgenstern, O., 2004. Theory of Games and 
Economic Behavior. Princeton: Princeton University.

\vskip 2mm

Weidlich, W., 2000. Sociodynamics: A Systematic Approach to
Mathematical Modelling in the Social Sciences. 
Amsterdam: Harwood Academic.

\vskip 2mm

White, E.N., 1996. Stock Market Crashes and Speculative 
Manias. Cheltenham: Edward Elgar.

\vskip 2mm

Wilson, K. G., 1979. Problems in physics with many scales of 
length. Scientific American 241, 158-179.

\vskip 2mm

Yukalov, V.I., 1990. Self-similar approximations for strongly 
interacting systems.  Physica A 167, 833-860. 

\vskip 2mm

Yukalov, V.I., 1991. Method of self-similar approximations. 
Journal of Mathematical Physics 32, 1235-1239. 

\vskip 2mm

Yukalov, V.I., 1992. Stability conditions for method of 
self-similar approximations. Journal of Mathematical Physics 33, 
3994-4001.

\vskip 2mm

Yukalov, V.I., 2000. Self-similar extrapolation of asymptotic 
series and forecasting for time series. Modern Physics Letters 
B 14, 791-800.

\vskip 2mm

Yukalov, V.I., 2001. Self-similar approach to market analysis. 
European Physical Journal B 20, 609-617.

\vskip 2mm

Yukalov, V.I., 2002. Stochastic instability of quasi-isolated 
systems. Physical Review E 65, 056118-11.

\vskip 2mm

Yukalov, V.I., 2003. Expansion exponents for non-equilibrium 
systems. Physica A 320, 149-168. 

\vskip 2mm

Yukalov, V.I., Gluzman, S., 1997. Self-similar bootstrap of 
divergent series. Physical Review E 55, 6552-6565.

\vskip 2mm

Yukalov, V.I., Gluzman, S., 1998. Self-similar exponential 
approximants. Physical Review E 58, 1359-1382.

\vskip 2mm

Yukalov, V.I., Shumovsky, A.S., 1990. Lectures on Phase 
Transitions. Singapore: World Scientific.

\vskip 2mm

Yukalov, V.I., Yukalova, E.P., 1993. Self-similar approximations 
and evolution equations. Nuovo Cimento B 108, 1017-1042. 

\vskip 2mm

Yukalov, V.I., Yukalova, E.P., 1996. Temporal dynamics in 
perturbation theory. Physica A 225, 336-362. 

\vskip 2mm

Yukalov, V.I., Yukalova, E.P., 1999. Self-similar perturbation 
theory. Annals of Physics 277, 219-254. 

\vskip 2mm

Yukalov, V.I., Yukalova, E.P., 2002. Self-similar structures 
and fractal transforms in approximation theory. 
Chaos Solitons and Fractals 14, 839-861. 

\vskip 2mm

Zhou, W.-X., Sornette, D., 2006. Self-fulfilling Ising model of 
financial markets.  European Physical Journal B. 
DOI: 10.1140/epjb/e2006-00391-6.
}

\newpage

\begin{center}
{\Large{\bf Table Captions}}
\end{center}

{\bf Table 1}. Limiting cases of markets. The market friction 
coefficient $\bt$ is assumed to be always negative, $\bt<0$, and 
the nonlinear trend followers are supposed to be characterized by 
positive feedback, $B>0$.

\vskip 5mm

{\bf Table 2}. Different types of markets with  individual and 
collective mean-reverting agents $(\al<0,\; A<0)$. In this case, the 
equilibrium point is also present, corresponding to the co-integrated 
observed and fundamental prices. One or two limit ``business'' cycles
appear and coexist with the equilibrium point for sufficiently 
unregulated markets.

\vskip 5mm

{\bf Table 3}. Different types of markets with individual 
speculative and collective mean-reverting agents $(\al>0,\; A<0)$. 
For low and intermediate uncertainty, only fixed points exist, 
the equilibrium point coexisting with a positive and a negative 
conventions. The corresponding stochastic dynamics is co-integrated 
with the price corresponding to one of these fixed points, but not 
for ever, as the coexistence between the three fixed points implies 
that the dynamics jumps spontaneously from one regime to another 
at random times. For large uncertainty and for weak regulations, 
limit business cycles appear in addition, competing with the other 
fixed points.

\vskip 5mm

{\bf Table 4}. Different types of markets with individual 
mean-reverting and collective speculative agents $(\al<0,\; A>0)$.
The larger uncertainty and weaker regulations introduce more 
coexisting fixed points and limit cycles.

\newpage

\begin{center}
{\Large{\bf Figure Captions}}
\end{center}

{\bf Fig. 1}. Phase portrait of a strictly regulated market with 
low uncertainty and individual as well as collective mean-reverting 
agents $(\al<0,\; A<0)$. The parameters are: $\al=-1$, $A=-10$, 
$\bt=-1$, $B=1$, $\mu=1$, $\lbd=1$. One stable fixed point $\{ 0,0\}$.

\vskip 5mm

{\bf Fig. 2}. Phase portrait of a weakly regulated market with 
low uncertainty and individual as well as collective mean-reverting 
agents The parameters are: $\al=-1$, $A=-10$, $\bt=-1$, $B=1$, 
$\mu=1$, $\lbd=2$. Stable fixed point $\{ 0,0\}$ and a limit cycle.

\vskip 5mm

{\bf Fig. 3}. Phase portrait of a market with low uncertainty 
and individual as well as collective mean-reverting agents, with 
increasing deregulation. The parameters are: $\al=-1$, $A=-10$, 
$\bt=-1$, $B=1$, $\mu=1$, $\lbd=3$. Stable fixed point $\{ 0,0\}$ 
and a growing limit cycle.

\vskip 5mm

{\bf Fig. 4}. Phase portrait of a strictly regulated market with 
large uncertainty and individual as well as collective mean-reverting 
agents. The parameters are: $\al=-1$, $A=-10$, $\bt=-1$, $B=1$, $\mu=2$, 
$\lbd=1$. One stable fixed point $\{ 0,0\}$.

\vskip 5mm

{\bf Fig. 5}. Phase portrait of an uncertain market with
individual as well as collective mean-reverting agents, with slightly 
increased liberalization. The parameters are: $\al=-1$, $A=-10$, 
$\bt=-1$, $B=1$, $\mu=2$, $\lbd=2$. Stable fixed point $\{ 0,0\}$ and 
a limit cycle.

\vskip 5mm

{\bf Fig. 6}. Phase portrait of an uncertain market with individual 
as well as collective mean-reverting agents, with intermediate 
liberalization. The parameters are: $\al=-1$, $A=-10$, $\bt=-1$, 
$B=1$, $\mu=2$, $\lbd=4$. Stable point $\{ 0,0\}$ and two limit cycles, 
internal and external.

\vskip 5mm

{\bf Fig. 7}. Phase portrait of an uncertain market with individual 
as well as collective mean-reverting agents, with strong liberalization.
The parameters are: $\al=-1$, $A=-10$, $\bt=-1$, $B=1$, $\mu=2$, 
$\lbd=5$. Stable point $\{ 0,0\}$ and one external limit cycle.

\vskip 5mm

{\bf Fig. 8}. Phase portrait of an uncertain strictly regulated 
market with individual speculative and collective mean-reverting 
agents. The parameters are: $\al=5$, $A=-1$, $\bt=-1$, $B=1$, $\mu=5$, 
$\lbd=1$. Two stable fixed points $\{ \pm s,0\}$.

\vskip 5mm

{\bf Fig. 9}. Phase portrait of an uncertain slightly deregulated 
market with individual speculative and collective mean-reverting 
agents. The parameters are: $\al=5$, $A=-1$, $\bt=-1$, $B=1$, $\mu=5$, 
$\lbd=2$. Two stable fixed points $\{ \pm s,0\}$ and two limit cycles 
around each of them.

\vskip 5mm

{\bf Fig. 10}. Phase portrait of an uncertain market with 
individual speculative and collective mean-reverting agents, and 
with intermediate level of liberalization. The parameters are: 
$\al=5$, $A=-1$, $\bt=-1$, $B=1$, $\mu=5$, $\lbd=3$. Two stable 
fixed points $\{ \pm s,0\}$ and one big stable limit cycle 
surrounding both points.

\vskip 5mm

{\bf Fig. 11}. Phase portrait of an uncertain strongly deregulated 
market with individual speculative and collective mean-reverting 
agents. The parameters are: $\al=5$, $A=-1$, $\bt=-1$, $B=1$, $\mu=5$, 
$\lbd=4$. Two stable fixed points $\{ \pm s,0\}$.

\vskip 5mm

{\bf Fig. 12}. Phase portrait of a strictly regulated market with 
low uncertainty, and with individual mean-reverting and collectively 
speculating agents. The parameters are: $\al=-1$, $A=1$, $\bt=-1$, 
$B=1$, $\mu=1$, $\lbd=1$. One stable fixed point $\{ 0,0\}$.

\vskip 5mm

{\bf Fig. 13}. Phase portrait of a market with low uncertainty, 
lower regulation, and with individual mean-reverting and 
collectively speculating agents. The parameters are: $\al=-1$, 
$A=1$, $\bt=-1$, $B=1$, $\mu=1$, $\lbd=2$. One stable fixed point 
$\{ 0,0\}$ and a stable limit cycle around it.

\vskip 5mm

{\bf Fig. 14}. Phase portrait of a market with low uncertainty, 
with increased liberalization and with individual mean-reverting 
and collectively speculating agents. The parameters are: $\al=-1$, 
$A=1$, $\bt=-1$, $B=1$, $\mu=1$, $\lbd=4$. The stable fixed point 
$\{ 0,0\}$ and a stable limit cycle, which is increased and deformed. 

\vskip 5mm

{\bf Fig. 15}. Phase portrait of a strictly regulated market with 
intermediate uncertainty, and with individual mean-reverting and 
collectively speculating agents. The parameters are: $\al=-1$, 
$A=1$, $\bt=-1$, $B=1$, $\mu=2$, $\lbd=1$. Three stable fixed points 
$\{ 0,0\}$ and $\{ \pm s,0\}$. 

\vskip 5mm

{\bf Fig. 16}. Phase portrait of a weakly regulated market with 
intermediate uncertainty, and with individual mean-reverting and 
collectively speculating agents. The parameters are: $\al=-1$, 
$A=1$, $\bt=-1$, $B=1$, $\mu=2$, $\lbd=2$. Three stable fixed points, 
$\{ 0,0\}$ and $\{ \pm s,0\}$, and a stable limit cycle surrounding 
all of them.

\vskip 5mm

{\bf Fig. 17}. Phase portrait of a market with an intermediate 
regulation, with large uncertainty, and with individual 
mean-reverting and collectively speculating agents. The parameters 
are: $\al=-1$, $A=1$, $\bt=-1$, $B=1$, $\mu=3$, $\lbd=2$. Three 
stable fixed points, $\{ 0,0\}$ and $\{\pm s,0\}$, and two stable 
limit cycles surrounding the points $\{ \pm s,0\}$.

\vskip 5mm

{\bf Fig. 18}.  Phase portrait of a weakly regulated market 
with large uncertainty, and with individual mean-reverting and 
collectively speculating agents. The parameters are: $\al=-1$, 
$A=1$, $\bt=-1$, $B=1$, $\mu=3$, $\lbd=3$. Three stable fixed points, 
$\{ 0,0\}$ and $\{\pm s,0\}$, with a stable limit cycle surrounding 
all of them.

\vskip 5mm

{\bf Fig. 19}. Influence of increasing volatility on the 
mispricing trajectory $x(t)$, with the initial conditions $x(0)=0$ 
and $y(0)=1$, for the case of a low-uncertainty liberal market with 
all mean-reverting agents. The corresponding phase portrait in the 
absence of stochasticity is given by Fig. 3. Switching on the 
stochasticity yields the following trajectories: (a) $\sgm=0$; 
(b) $\sgm=0.1$; (c) $\sgm=0.5$; (d) $\sgm=5$. All other parameters 
are the same as in Fig. 3. 

\vskip 5mm

{\bf Fig. 20}.  Influence of stochasticity on the mispricing 
trajectory $x(t)$, with the initial conditions $x(0)=3$ and 
$y(0)=0$, for the uncertain slightly deregulated market with 
individual speculative and collective reverting agents. All market 
parameters are as in Fig. 9. The increasing stochasticity results 
in the following trajectories: (a) $\sgm=0$; (b) $\sgm=0.1$; 
(c) $\sgm=0.3$; (d) $\sgm=3$. 

\vskip 5mm

{\bf Fig. 21}. Influence of stochasticity on the strictly regulated 
market with intermediate uncertainty, whose phase portrait and market 
parameters are given by Fig. 15. Initial conditions are: $x(0)=1.5$ 
and $y(0)=0$. The volatility parameters are: (a) $\sgm=0$; 
(b) $\sgm=0.2$; (c) $\sgm=1$; (d) $\sgm=3$. 

\vskip 5mm

{\bf Fig. 22}. Influence of stochasticity on an uncertain market 
with an intermediate regulation. Initial conditions: $x(0)=5$ and 
$y(0)=0$. The related phase portrait and market parameters are from 
Fig. 17. The stochasticity strength is regulated by the parameter 
$\sgm$ in the following way: (a) $\sgm=0$; (b) $\sgm=1$; (c) $\sgm=3$; 
(d) $\sgm=5$. 

\vskip 5mm

{\bf Fig. 23}. Influence of stochasticity on a weakly regulated 
uncertain market represented by the phase portrait of Fig. 18. 
Initial conditions: $x(0)=1$ and $y(0)=0$. All market parameters 
are the same as in the latter figure. Here, the stochasticity 
parameters are: (a) $\sgm=0$; (b) $\sgm=3$. 

\vskip 5mm

{\bf Fig. 24}. Long-time behavior of the mispricing trajectories 
demonstrating the jumps between three trapping regions around the 
points $\{0,0\}$, $\{-s,0\}$, and $\{ s,0\}$. The market parameters 
are the same as in Fig. 15 for a strictly regulated market with 
intermediate uncertainty, and with individual reverting while 
collectively speculative agents. Initial conditions are: $x(0)=1.5$ 
and $y(0)=0$. The volatility parameter is $\sgm=0.9$. 

\vskip 5mm

{\bf Fig. 25}. Long-time behavior of the mispricing for the 
uncertain weakly regulated market with individual reverting and 
collectively speculative agents, characterized by the phase portrait 
of Fig. 17. All market parameters are the same as in this figure. 
Initial conditions are: $x(0)=5$ and $y(0)=0$. The volatility 
parameter is $\sgm=2$. The trajectory jumps between the trapping 
regions corresponding to the points $\{0,0\}$, $\{-s,0\}$, and 
$\{ s,0\}$, and the trapping regions related to the cycles $C_{-s}$ 
and $C_s$. 

\newpage 

\begin{center}

\begin{tabular}{|c|c|c|} \hline
Market                      &  Globally stable   & Locally stable  \\ \hline
super-certain $(\mu\ra 0)$      & $\{ 0,0\}$ $\al<0$ & \\
overregulated $(\lbd\ra 0)$ &                    & \\ \hline
super-certain $(\mu\ra 0)$      &                    & $\{ 0,0\}$ $\al<0$ \\
unregulated $(\lbd\ra\infty)$&                   &   \\ \hline
super-uncertain $(\mu\ra\infty)$  &$\{ 0,0\}$ $\al<0$, $A<0$&  $\{ 0,0\}$ $\al<0$, $A>0$\\
overregulated $(\lbd\ra 0)$ &$\{\pm s,0\}$ $\al>0$, $A<0$&   \\ \hline
super-uncertain $(\mu\ra\infty)$  &                    & $\{ 0,0\}$ $\al<0$ \\
unregulated $(\lbd\ra\infty)$&                   & $\{\pm s,0\}$ $\al>0$, $A<0$ \\ \hline
\end{tabular}

\vskip 1cm

\end{center}

{\parindent=0pt
Table 1: Limiting cases of markets. The market friction coefficient 
$\bt$ is assumed to be always negative, $\bt<0$, and the nonlinear 
trend followers are supposed to be characterized by positive feedback, 
$B>0$.}

\newpage

\begin{center}

\begin{tabular}{|c|c|c|} \hline
$\al<0$, $A<0$  & regulation & attractors \\ \hline
low uncertainty & rigid: $0<\lbd<\lbd_1$     & $\{ 0,0\}$ \\
$0<\mu<\mu_1$   & soft: $\lbd_1<\lbd<\infty$ & $\{ 0,0\}$ $C_0$ \\ \hline
high uncertainty& rigid: $0<\lbd<\lbd_1$     & $\{ 0,0\}$ \\
$\mu_1<\mu<\infty$& mid-rigid: $\lbd_1<\lbd<\lbd_2$& $\{ 0,0\}$ $C_0$ \\
                 & mid-soft: $\lbd_2<\lbd<\lbd_3$& $\{ 0,0\}$ $C_0$ $C_0'$\\
                 & soft: $\lbd_3<\lbd<\infty$& $\{ 0,0\}$ $C_0'$ \\ \hline
\end{tabular}

\vskip 1cm

\end{center}

{\parindent=0pt
Table 2: Different types of markets with individual and collective 
mean-reverting agents $(\al<0,\; A<0)$. In this case, the equilibrium 
point is also present, corresponding to the co-integrated observed and 
fundamental prices. One or two limit ``business'' cycles appear and 
coexist with the equilibrium point for sufficiently unregulated markets.}

\newpage

\begin{center}

\begin{tabular}{|c|c|c|c|} \hline
$\al>0$, $A<0$  & speculation & regulation & attractors \\ \hline
intermediate uncertainty &             &            &          \\
$\mu_c<\mu<\mu_1$& any: $0<\al<\infty$ & any: $0<\lbd<\infty$ & 
$\{\pm s,0\}$\\ \hline
large uncertainty & weak    & rigid: $0<\lbd<\lbd_1$ & $\{\pm s,0\}$ \\
$\mu_1<\mu<\infty$ &$0<\al<\al_c$ & mid-soft: $\lbd_1<\lbd<\lbd_2$& 
$\{\pm s,0\}$ $C_{\pm s}$\\
                   &             & soft: $\lbd_2<\lbd<\infty$& 
$\{\pm s,0\}$ \\ \hline
large uncertainty  & strong  & rigid: $0<\lbd<\lbd_1$ & $\{\pm s,0\}$ \\
$\mu_1<\mu<\infty$ & $\al_c<\al<\infty$ & mid-rigid: $\lbd_1<\lbd<\lbd_2$& 
$\{\pm s,0\}$ $C_{-s}$ $C_s$ \\
     &            & mid-soft: $\lbd_2<\lbd<\lbd_3$& $\{\pm s,0\}$ $C_{\pm s}$\\
     &            & soft: $\lbd_3<\lbd<\infty$&  $\{\pm s,0\}$ \\ \hline  
\end{tabular}

\vskip 1cm

\end{center}

{\parindent=0pt
Table 3: Different types of markets with individual speculative 
and collective mean-reverting agents $(\al>0,\; A<0)$. For low and 
intermediate uncertainty, only fixed points exist, the equilibrium 
point coexisting with a positive and a negative conventions. The 
corresponding stochastic dynamics is co-integrated with the price 
corresponding to one of these fixed points, but not for ever, as the 
coexistence between the three fixed points implies that the dynamics 
jumps spontaneously from one regime to another at random times. For 
large uncertainty and for weak regulations, limit business cycles 
appear in addition, competing with the other fixed points.}

\newpage

\begin{center}

\begin{tabular}{|c|c|c|} \hline
$\al<0$, $A>0$  & regulation & attractors \\ \hline
low uncertainty & rigid: $0<\lbd<\lbd_1$ & $\{ 0,0\}$ \\
$0<\mu<\mu_c$   & soft: $\lbd_1<\lbd<\infty$ & $\{ 0,0\}$ $C_0$ \\ \hline
intermediate uncertainty& rigid: $0<\lbd<\lbd_1$ & $\{ 0,0\}$ $\{\pm s,0\}$ \\
$\mu_c<\mu<\mu_1$ & soft: $\lbd_1<\lbd<\infty$ & $\{ 0,0\}$ $\{\pm s,0\}$ 
$C_{0,\pm s}$\\ \hline
large uncertainty  & rigid: $0<\lbd<\lbd_1$ & $\{ 0,0\}$ $\{\pm s,0\}$ \\
$\mu_1<\mu<\infty$ & mid-soft: $\lbd_1<\lbd<\lbd_2$& $\{ 0,0\}$ $\{\pm s,0\}$ 
$C_{-s}$ $C_s$ \\
       & soft: $\lbd_2<\lbd<\infty$ & $\{ 0,0\}$ $\{\pm s,0\}$ $C_{0,\pm s}$\\ \hline
\end{tabular}

\vskip 1cm

\end{center}

{\parindent=0pt
Table 4: Different types of markets with individual mean-reverting 
and collective speculative agents $(\al<0,\;  A>0)$. The larger 
uncertainty and weaker regulations introduce more coexisting fixed 
points and limit cycles.}

\newpage

\begin{figure}[h]
\centerline{\psfig{file=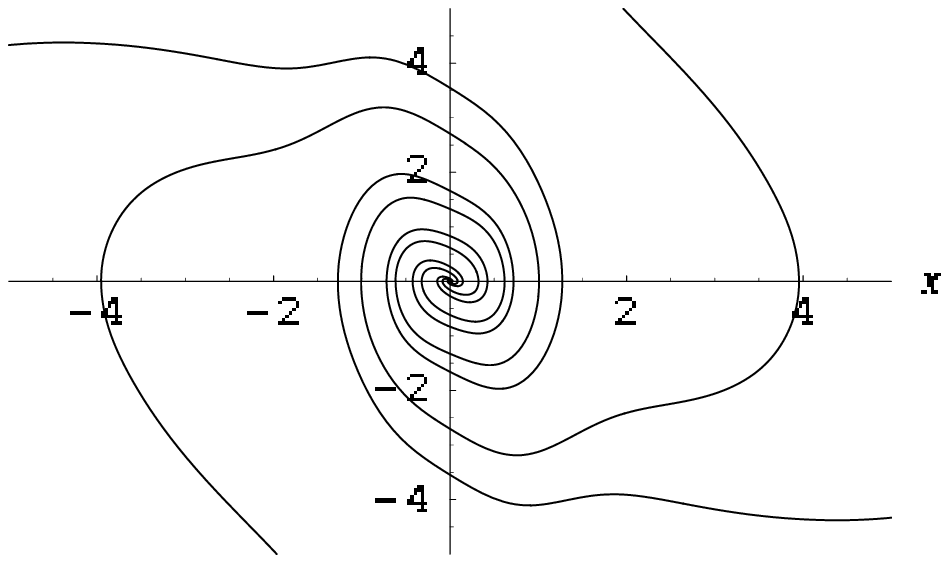,angle=0,width=16cm}}
\vskip 1cm
\caption{Phase portrait of a strictly regulated market with low
uncertainty and individual as well as collective mean-reverting agents 
$(\al<0,\; A<0)$. The parameters are: $\al=-1$, $A=-10$, $\bt=-1$, $B=1$, 
$\mu=1$, $\lbd=1$. One stable fixed point $\{ 0,0\}$.
}
\label{fig:Fig.1}
\end{figure}

\newpage

\begin{figure}[h]
\centerline{\psfig{file=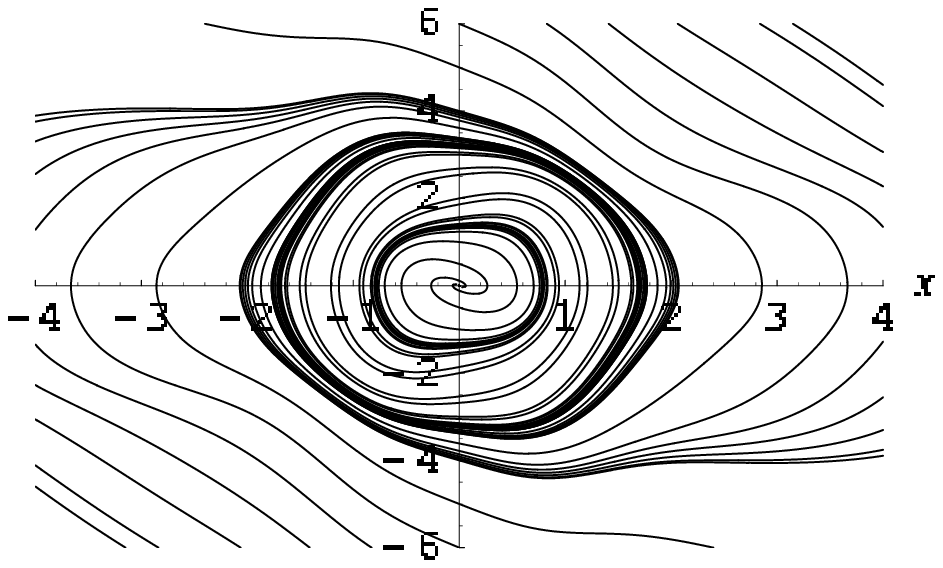,angle=0,width=16cm}}
\vskip 1cm
\caption{Phase portrait of a weakly regulated market with low uncertainty 
and individual as well as collective mean-reverting agents The parameters 
are: $\al=-1$, $A=-10$, $\bt=-1$, $B=1$, $\mu=1$, $\lbd=2$. Stable fixed 
point $\{ 0,0\}$ and a limit cycle.
}
\label{fig:Fig.2}
\end{figure}

\newpage

\begin{figure}[h]
\centerline{\psfig{file=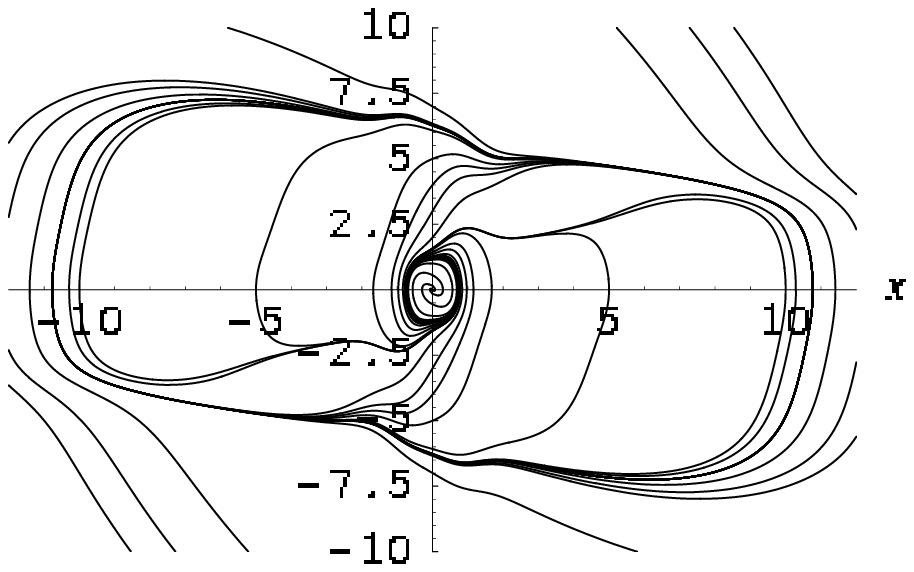,angle=0,width=16cm}}
\vskip 1cm
\caption{Phase portrait of a market with low uncertainty
and individual as well as collective mean-reverting agents, with 
increasing deregulation. The parameters are: $\al=-1$, $A=-10$, $\bt=-1$, 
$B=1$, $\mu=1$, $\lbd=3$. Stable fixed point $\{ 0,0\}$ and a growing 
limit cycle.
}
\label{fig:Fig.3}
\end{figure}

\newpage

\begin{figure}[h]
\centerline{\psfig{file=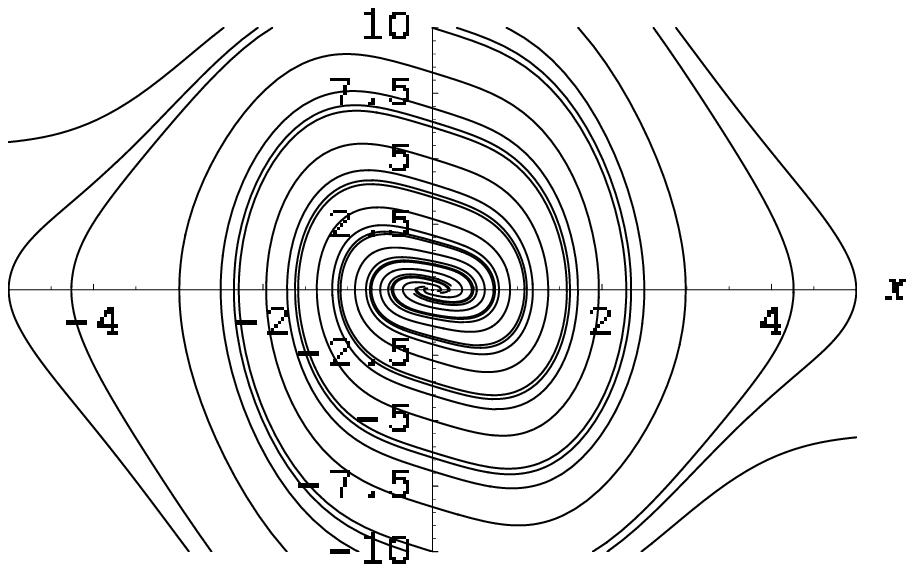,angle=0,width=16cm}}
\vskip 1cm
\caption{Phase portrait of a strictly regulated market with large 
uncertainty and individual as well as collective mean-reverting agents.
The parameters are: $\al=-1$, $A=-10$, $\bt=-1$, $B=1$, $\mu=2$, $\lbd=1$.
One stable fixed point $\{ 0,0\}$.
}
\label{fig:Fig.4}
\end{figure}

\newpage

\begin{figure}[h]
\centerline{\psfig{file=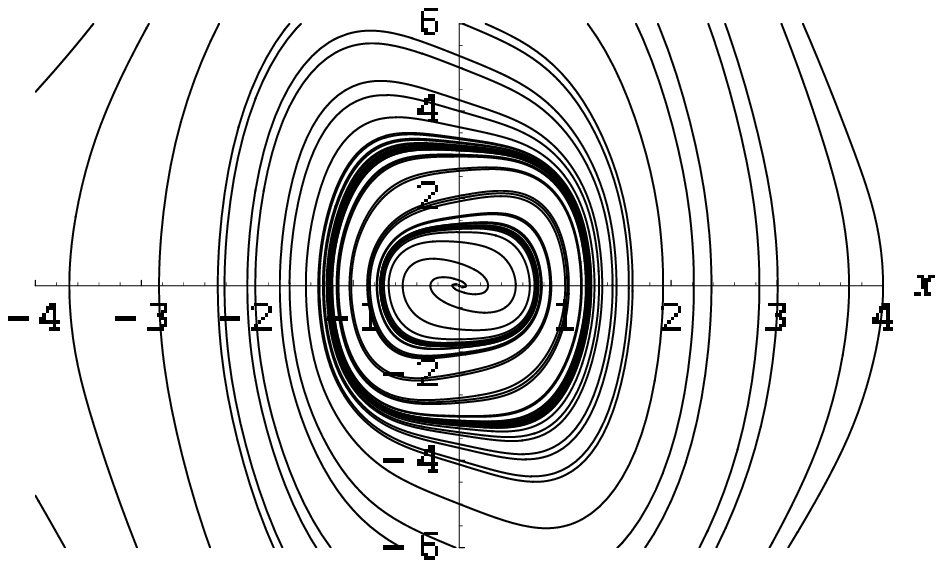,angle=0,width=16cm}}
\vskip 1cm
\caption{Phase portrait of an uncertain market with individual as 
well as collective mean-reverting agents, with slightly increased 
liberalization. The parameters are: $\al=-1$, $A=-10$, $\bt=-1$, 
$B=1$, $\mu=2$, $\lbd=2$. Stable fixed point $\{ 0,0\}$ and a limit 
cycle.
}
\label{fig:Fig.5}
\end{figure}

\newpage

\begin{figure}[h]
\centerline{\psfig{file=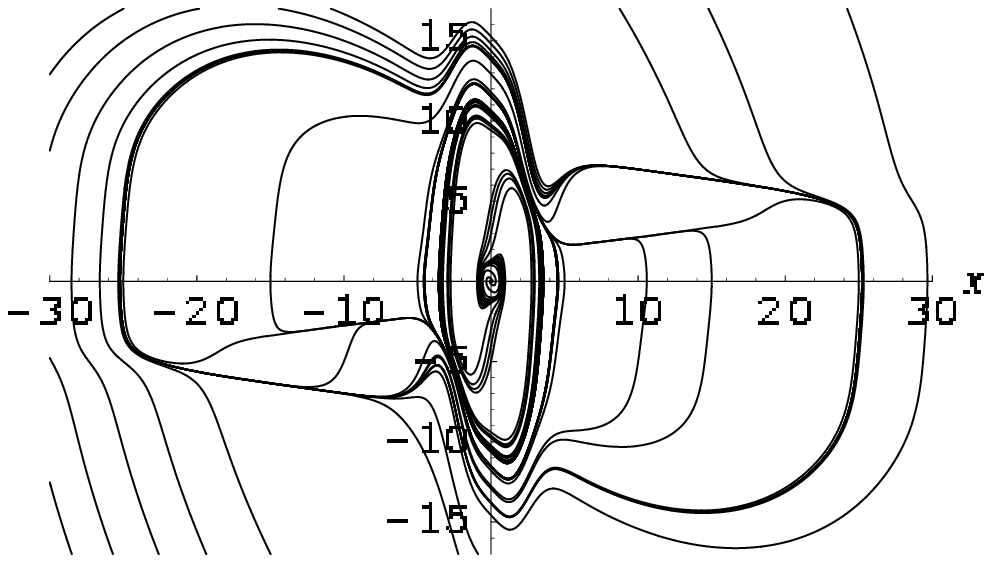,angle=0,width=16cm}}
\vskip 1cm
\caption{Phase portrait of an uncertain market with individual 
as well as collective mean-reverting agents, with intermediate 
liberalization. The parameters are: $\al=-1$, $A=-10$, $\bt=-1$, 
$B=1$, $\mu=2$, $\lbd=4$. Stable point $\{ 0,0\}$ and two limit 
cycles, internal and external.
}
\label{fig:Fig.6}
\end{figure}

\newpage

\begin{figure}[h]
\centerline{\psfig{file=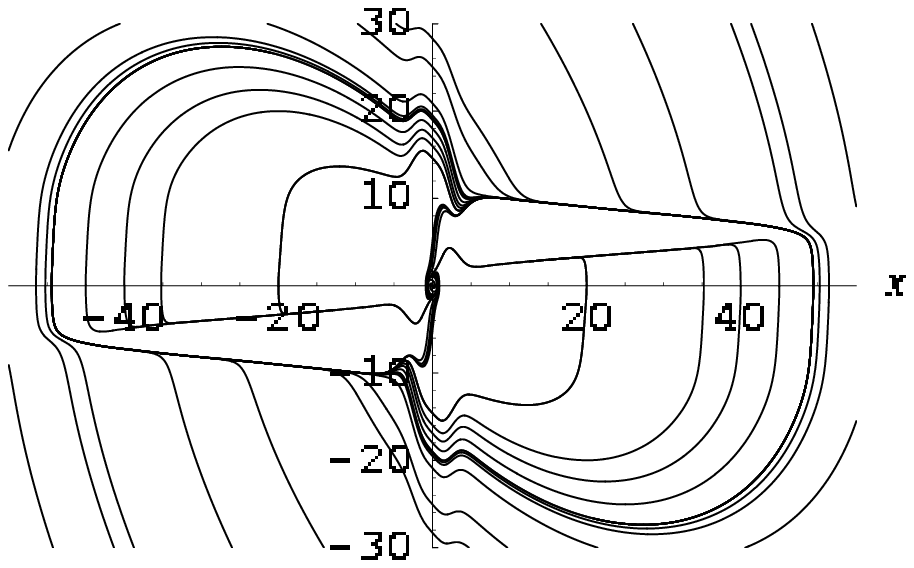,angle=0,width=16cm}}
\vskip 1cm
\caption{Phase portrait of an uncertain market with
individual as well as collective mean-reverting agents, with strong 
liberalization. The parameters are: $\al=-1$, $A=-10$, $\bt=-1$, $B=1$, 
$\mu=2$, $\lbd=5$. Stable point $\{ 0,0\}$ and one external limit cycle.
}
\label{fig:Fig.7}
\end{figure}

\newpage

\begin{figure}[h]
\centerline{\psfig{file=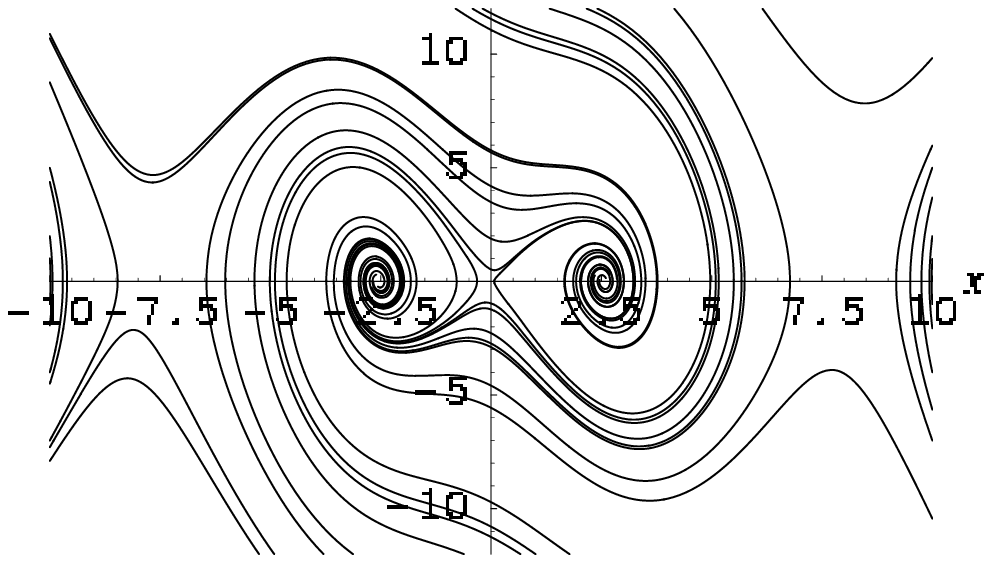,angle=0,width=16cm}}
\vskip 1cm
\caption{Phase portrait of an uncertain strictly regulated market 
with individual speculative and collective mean-reverting agents. The
parameters are: $\al=5$, $A=-1$, $\bt=-1$, $B=1$, $\mu=5$, $\lbd=1$.
Two stable fixed points $\{ \pm s,0\}$.
}
\label{fig:Fig.8}
\end{figure}

\newpage

\begin{figure}[h]
\centerline{\psfig{file=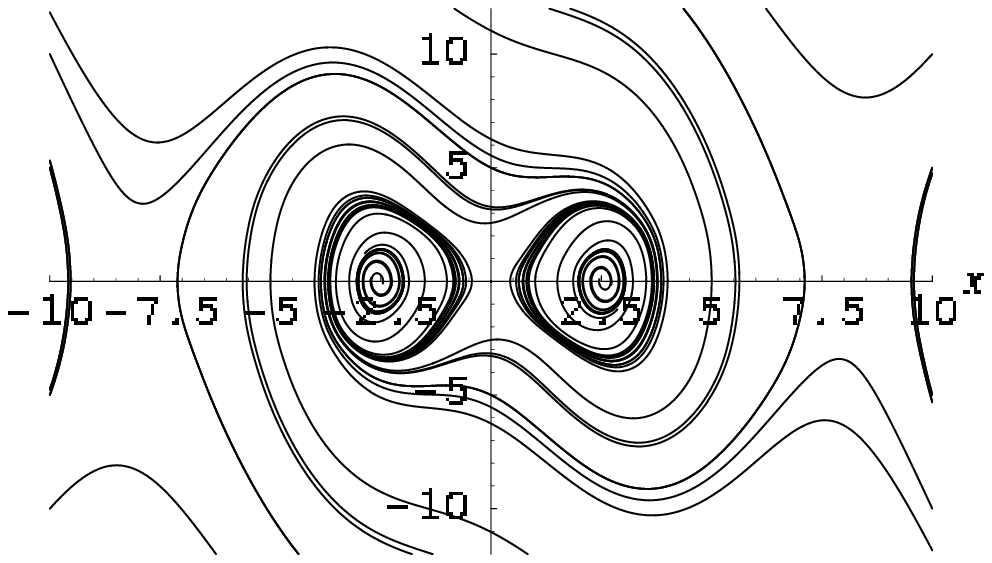,angle=0,width=16cm}}
\vskip 1cm
\caption{Phase portrait of an uncertain slightly deregulated market 
with individual speculative and collective mean-reverting agents.
The parameters are: $\al=5$, $A=-1$, $\bt=-1$, $B=1$, $\mu=5$, $\lbd=2$.
Two stable fixed points $\{ \pm s,0\}$ and two limit cycles around each 
of them.
}
\label{fig:Fig.9}
\end{figure}

\newpage

\begin{figure}[h]
\centerline{\psfig{file=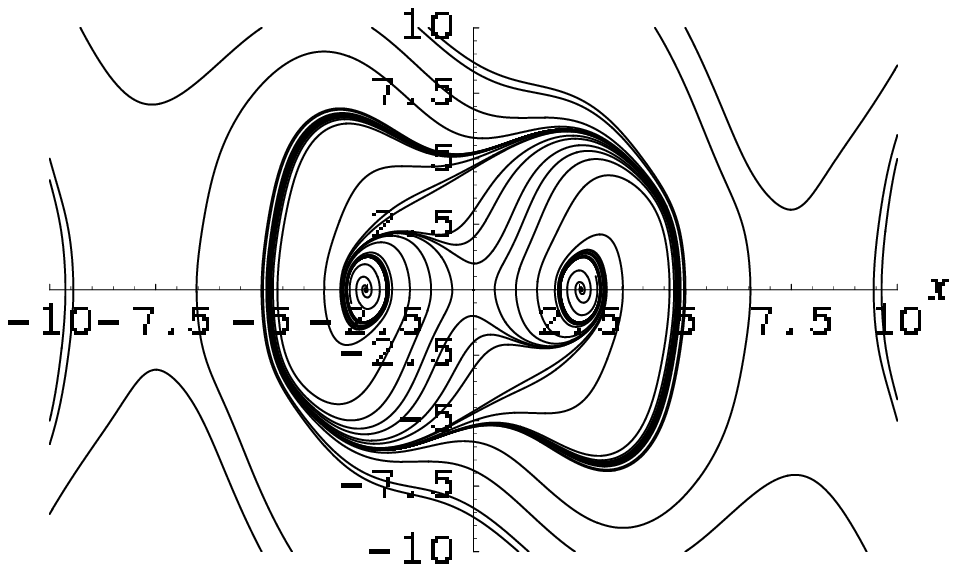,angle=0,width=16cm}}
\vskip 1cm
\caption{Phase portrait of an uncertain market with 
individual speculative and collective mean-reverting agents, and with
intermediate level of liberalization. The parameters are: $\al=5$, 
$A=-1$, $\bt=-1$, $B=1$, $\mu=5$, $\lbd=3$. Two stable fixed points 
$\{ \pm s,0\}$ and one big stable limit cycle surrounding both points.
}
\label{fig:Fig.10}
\end{figure}

\newpage

\begin{figure}[h]
\centerline{\psfig{file=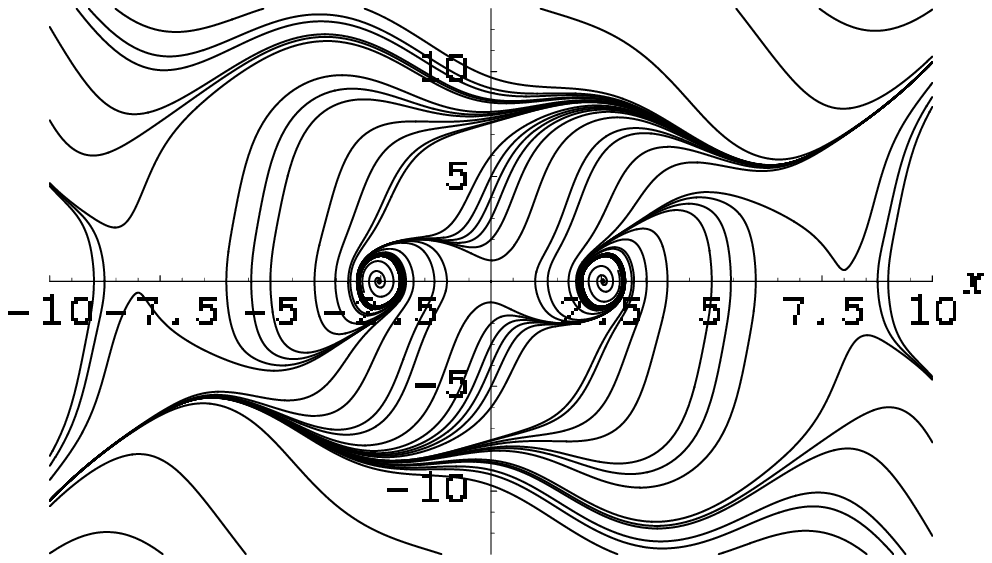,angle=0,width=16cm}}
\vskip 1cm
\caption{Phase portrait of an uncertain strongly deregulated market 
with individual speculative and collective mean-reverting agents.
The parameters are: $\al=5$, $A=-1$, $\bt=-1$, $B=1$, $\mu=5$, $\lbd=4$.
Two stable fixed points $\{ \pm s,0\}$.
}
\label{fig:Fig.11}
\end{figure}

\newpage

\begin{figure}[h]
\centerline{\psfig{file=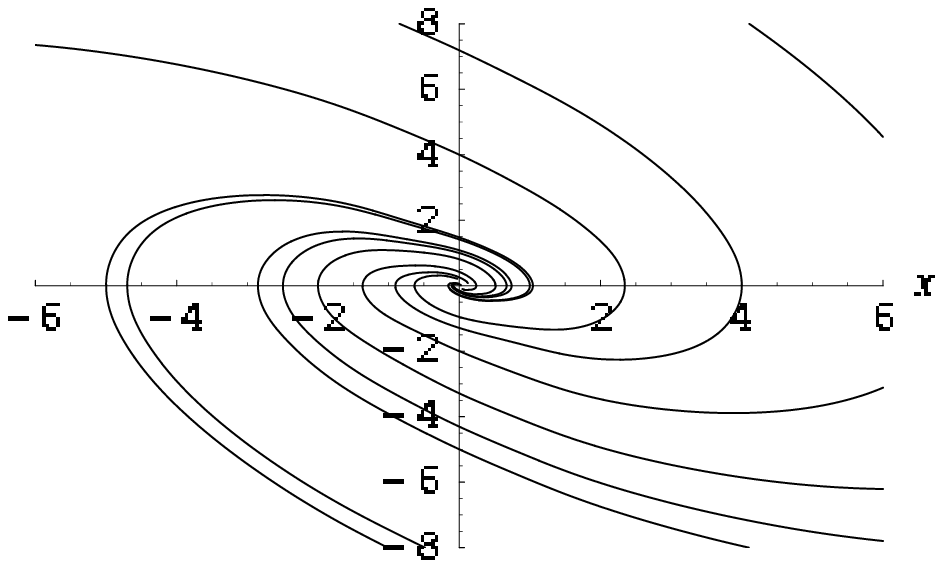,angle=0,width=16cm}}
\vskip 1cm
\caption{Phase portrait of a strictly regulated market with low 
uncertainty, and with individual mean-reverting and collectively 
speculating agents. The parameters are: $\al=-1$, $A=1$, $\bt=-1$, 
$B=1$, $\mu=1$, $\lbd=1$. One stable fixed point $\{ 0,0\}$.
}
\label{fig:Fig.12}
\end{figure}

\newpage

\begin{figure}[h]
\centerline{\psfig{file=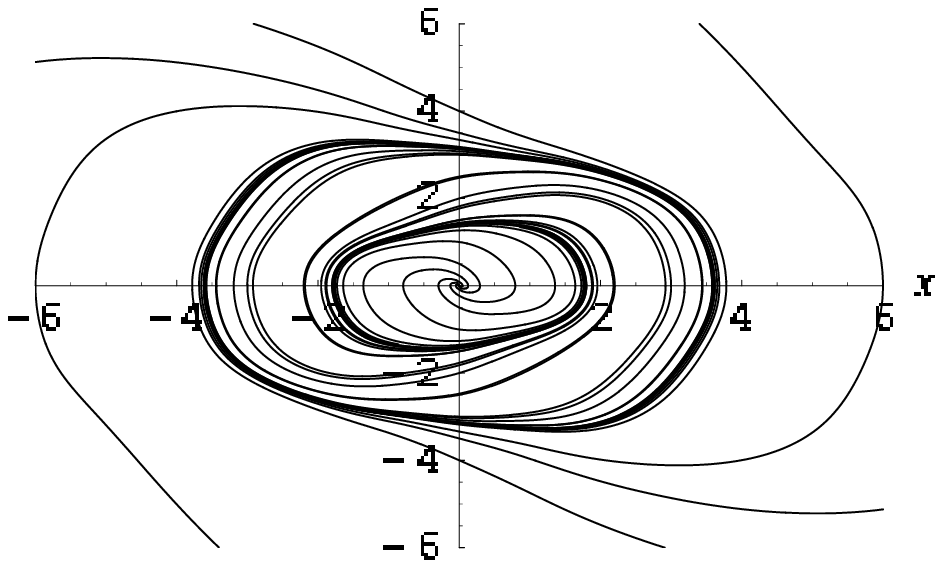,angle=0,width=16cm}}
\vskip 1cm
\caption{Phase portrait of a market with low uncertainty, lower 
regulation, and with individual mean-reverting and collectively 
speculating agents. The parameters are: $\al=-1$, $A=1$, $\bt=-1$, 
$B=1$, $\mu=1$, $\lbd=2$. One stable fixed point $\{ 0,0\}$ and a 
stable limit cycle around it.
}
\label{fig:Fig.13}
\end{figure}

\newpage

\begin{figure}[h]
\centerline{\psfig{file=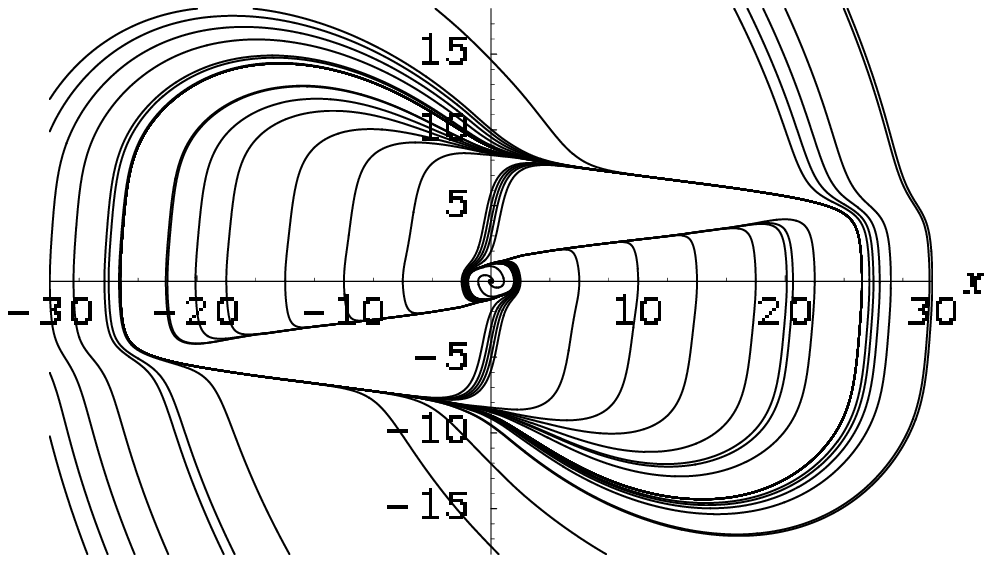,angle=0,width=16cm}}
\vskip 1cm
\caption{Phase portrait of a market with low uncertainty, with 
increased liberalization and with individual mean-reverting and 
collectively speculating agents. The parameters are: $\al=-1$, $A=1$, 
$\bt=-1$, $B=1$, $\mu=1$, $\lbd=4$. The stable fixed point $\{ 0,0\}$ 
and a stable limit cycle, which is increased and deformed. 
}
\label{fig:Fig.14}
\end{figure}

\newpage

\begin{figure}[h]
\centerline{\psfig{file=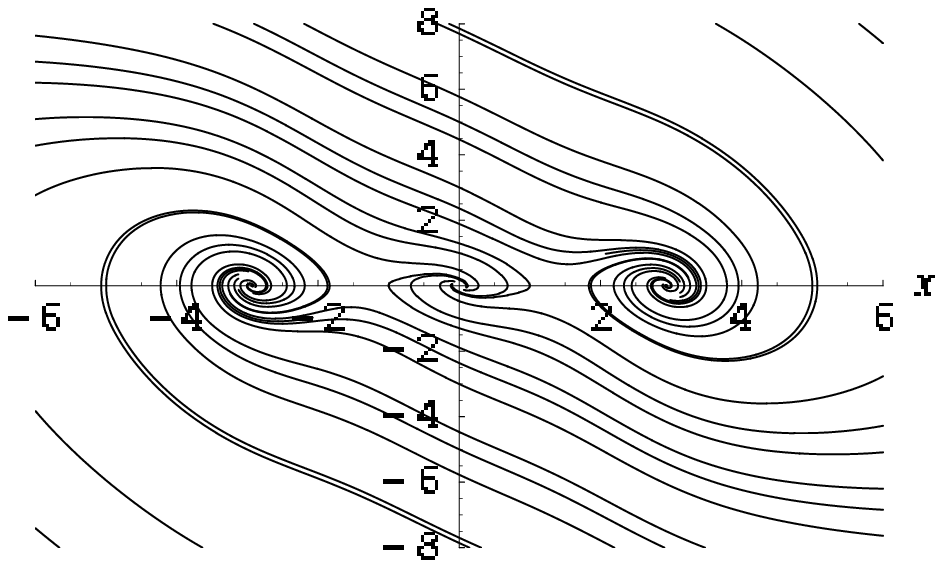,angle=0,width=16cm}}
\vskip 1cm
\caption{Phase portrait of a strictly regulated market with 
intermediate uncertainty, and with individual mean-reverting and 
collectively speculating agents. The parameters are: $\al=-1$, $A=1$, 
$\bt=-1$, $B=1$, $\mu=2$, $\lbd=1$. Three stable fixed points $\{ 0,0\}$ 
and $\{ \pm s,0\}$.
}
\label{fig:Fig.15}
\end{figure}

\newpage

\begin{figure}[h]
\centerline{\psfig{file=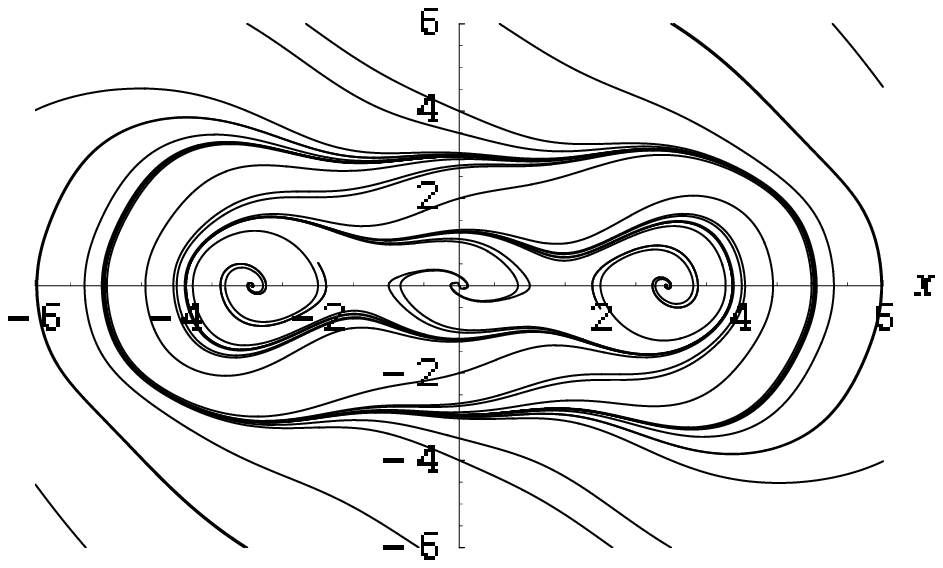,angle=0,width=16cm}}
\vskip 1cm
\caption{Phase portrait of a weakly regulated market with 
intermediate uncertainty, and with individual mean-reverting and 
collectively speculating agents. The parameters are: $\al=-1$, $A=1$, 
$\bt=-1$, $B=1$, $\mu=2$, $\lbd=2$. Three stable fixed points, 
$\{ 0,0\}$ and $\{ \pm s,0\}$, and a stable limit cycle surrounding 
all of them.
}
\label{fig:Fig.16}
\end{figure}

\newpage

\begin{figure}[h]
\centerline{\psfig{file=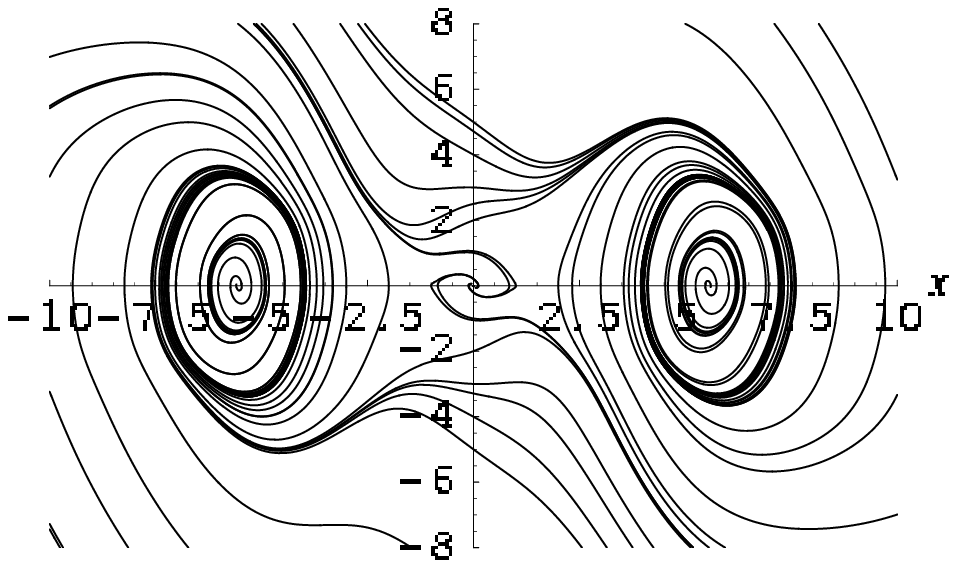,angle=0,width=16cm}}
\vskip 1cm
\caption{Phase portrait of a market with an intermediate regulation,
with large uncertainty, and with individual mean-reverting and 
collectively speculating agents. The parameters are: $\al=-1$, $A=1$, 
$\bt=-1$, $B=1$, $\mu=3$, $\lbd=2$. Three stable fixed points, $\{ 0,0\}$ 
and $\{\pm s,0\}$, and two stable limit cycles surrounding the points 
$\{ \pm s,0\}$.
}
\label{fig:Fig.17}
\end{figure}

\newpage

\begin{figure}[h]
\centerline{\psfig{file=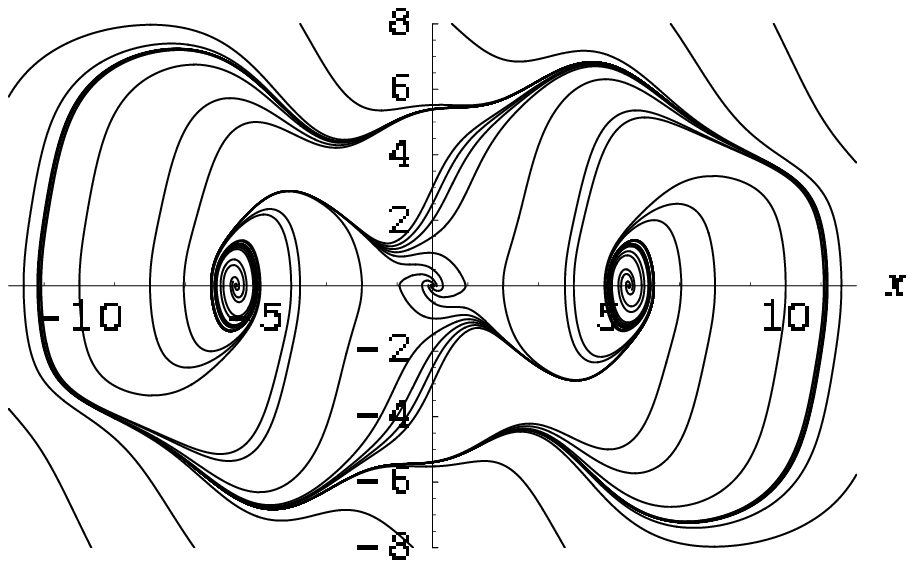,angle=0,width=16cm}}
\vskip 1cm
\caption{Phase portrait of a weakly regulated market with large 
uncertainty, and with individual mean-reverting and collectively 
speculating agents. The parameters are: $\al=-1$, $A=1$, $\bt=-1$, 
$B=1$, $\mu=3$, $\lbd=3$. Three stable fixed points, $\{ 0,0\}$ and 
$\{\pm s,0\}$, with a stable limit cycle surrounding all of them.
}
\label{fig:Fig.18}
\end{figure}

\newpage

\begin{figure}[h]
\centerline{\psfig{file=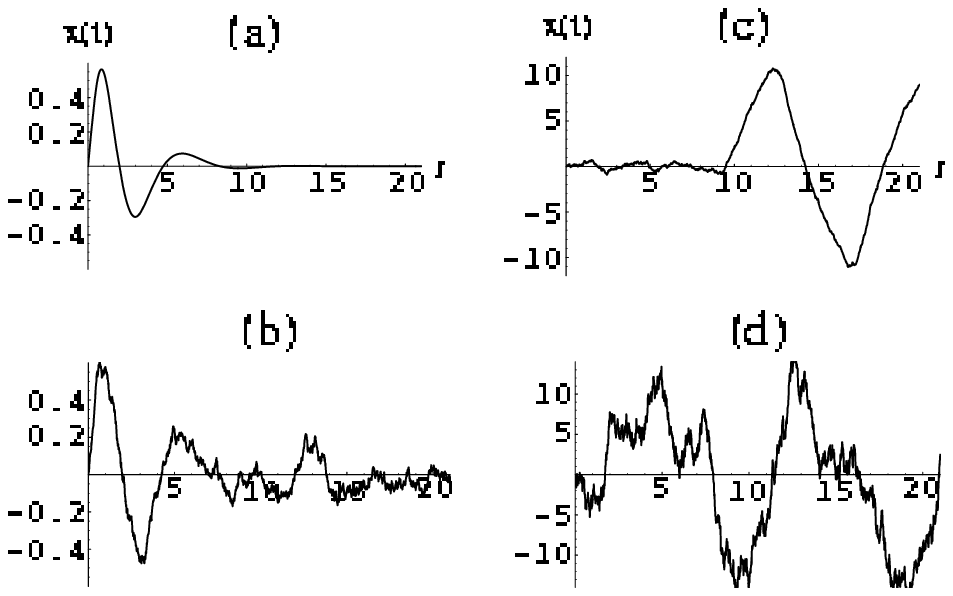,angle=0,width=16.5cm}}
\vskip 1cm
\caption{Influence of increasing volatility on the 
mispricing trajectory $x(t)$, with the initial conditions $x(0)=0$ 
and $y(0)=1$, for the case of a low-uncertainty liberal market with 
all mean-reverting agents. The corresponding phase portrait in the 
absence of stochasticity is given by Fig. 3. Switching on the 
stochasticity yields the following trajectories: (a) $\sgm=0$; 
(b) $\sgm=0.1$; (c) $\sgm=0.5$; (d) $\sgm=5$. All other parameters 
are the same as in Fig. 3. 
}
\label{fig:Fig.19}
\end{figure}

\newpage

\begin{figure}[h]
\centerline{\psfig{file=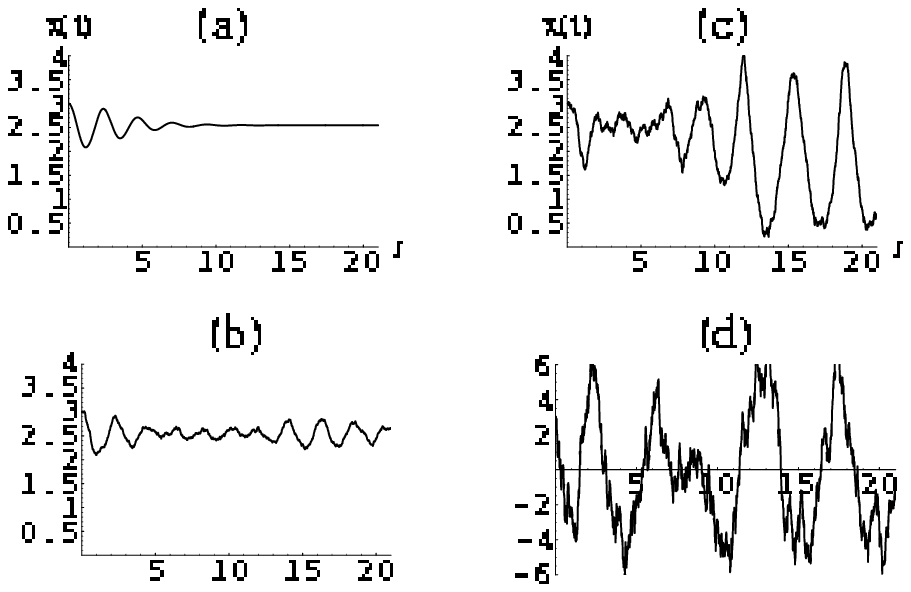,angle=0,width=16.5cm}}
\vskip 1cm
\caption{Influence of stochasticity on the mispricing trajectory 
$x(t)$, with the initial conditions $x(0)=3$ and $y(0)=0$, for the 
uncertain slightly deregulated market with individual speculative and 
collective reverting agents. All market parameters are as in Fig. 9. 
The increasing stochasticity results in the following trajectories: 
(a) $\sgm=0$; (b) $\sgm=0.1$; (c) $\sgm=0.3$; (d) $\sgm=3$.
}
\label{fig:Fig.20}
\end{figure}

\newpage

\begin{figure}[h]
\centerline{\psfig{file=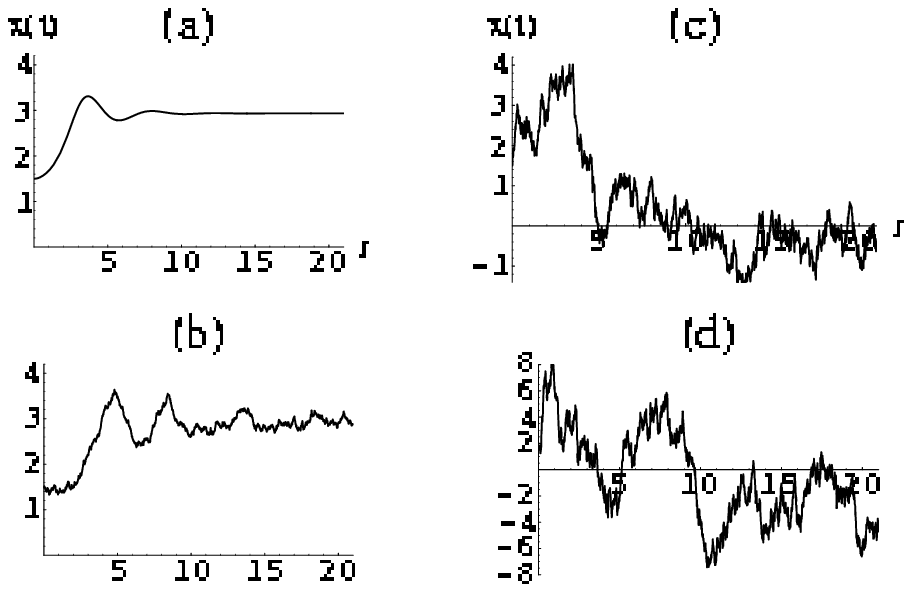,angle=0,width=16.5cm}}
\vskip 1cm
\caption{Influence of stochasticity on the strictly regulated market 
with intermediate uncertainty, whose phase portrait and market 
parameters are given by Fig. 15. Initial conditions are: $x(0)=1.5$ 
and $y(0)=0$. The volatility parameters are: (a) $\sgm=0$; (b) $\sgm=0.2$; 
(c) $\sgm=1$; (d) $\sgm=3$.
}
\label{fig:Fig.21}
\end{figure}

\newpage

\begin{figure}[h]
\centerline{\psfig{file=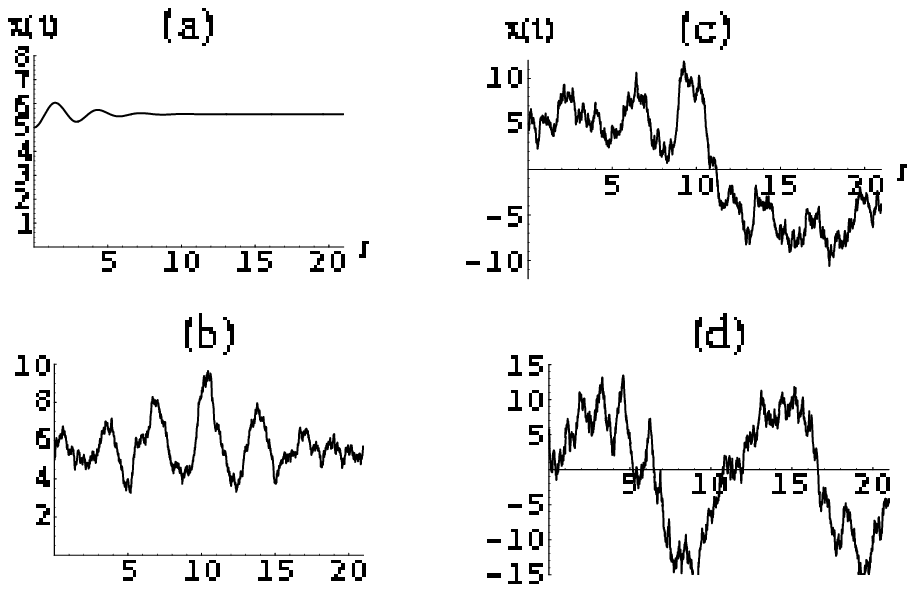,angle=0,width=16.5cm}}
\vskip 1cm
\caption{Influence of stochasticity on an uncertain market with an 
intermediate regulation. Initial conditions: $x(0)=5$ and $y(0)=0$. 
The related phase portrait and market parameters are from Fig. 17. 
The stochasticity strength is regulated by the parameter $\sgm$ in the 
following way: (a) $\sgm=0$; (b) $\sgm=1$; (c) $\sgm=3$; (d) $\sgm=5$.
}
\label{fig:Fig.22}
\end{figure}

\newpage

\begin{figure}[h]
\centerline{\psfig{file=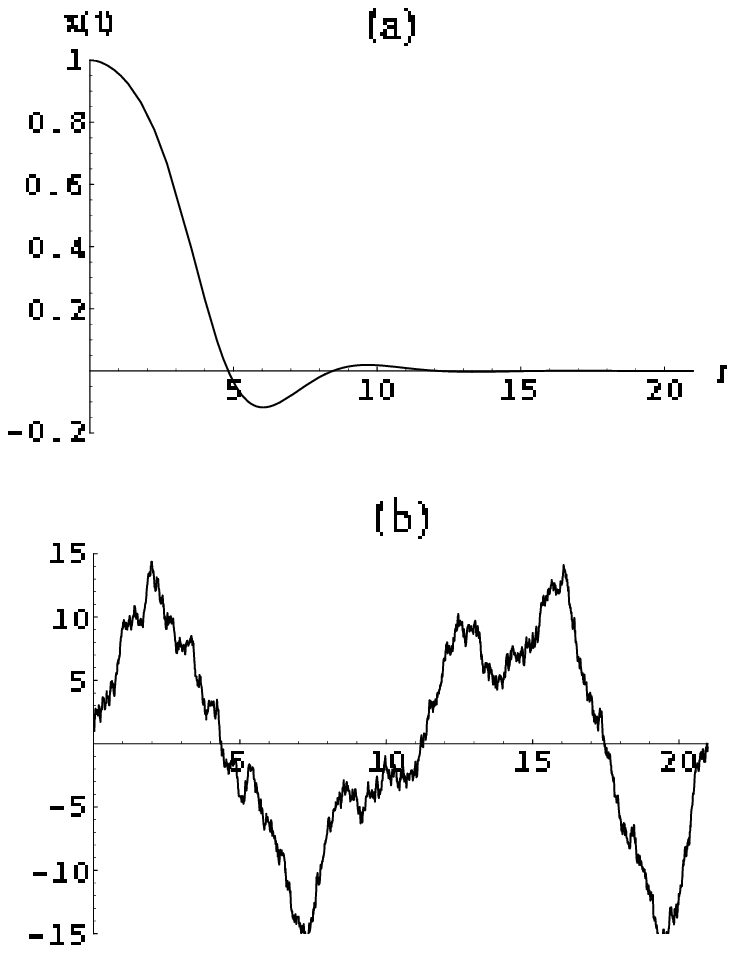,angle=0,width=12cm}}
\vskip 1cm
\caption{Influence of stochasticity on a weakly regulated uncertain 
market represented by the phase portrait of Fig. 18. Initial 
conditions: $x(0)=1$ and $y(0)=0$. All market parameters are the 
same as in the latter figure. Here, the stochasticity parameters 
are: (a) $\sgm=0$; (b) $\sgm=3$.
}
\label{fig:Fig.23}
\end{figure}

\newpage

\begin{figure}[h]
\centerline{\psfig{file=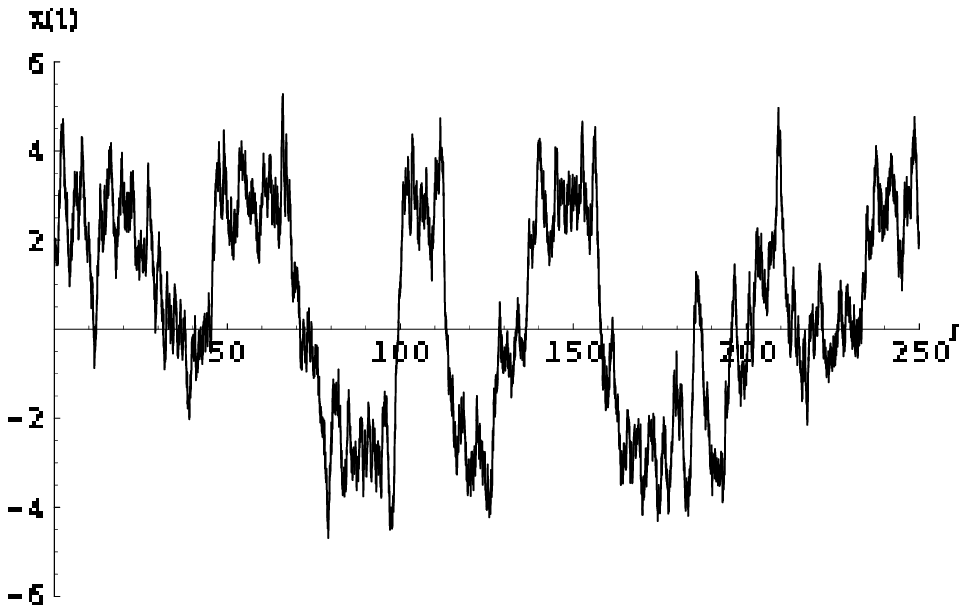,angle=0,width=12cm}}
\vskip 1cm
\caption{Long-time behavior of the mispricing trajectories 
demonstrating the jumps between three trapping regions around the 
points $\{0,0\}$, $\{-s,0\}$, and $\{ s,0\}$. The market parameters 
are the same as in Fig. 15 for a strictly regulated market with 
intermediate uncertainty, and with individual reverting while 
collectively speculative agents. Initial conditions are: $x(0)=1.5$ 
and $y(0)=0$. The volatility parameter is $\sgm=0.9$. 
}
\label{fig:Fig.24}
\end{figure}

\newpage

\begin{figure}[h]
\centerline{\psfig{file=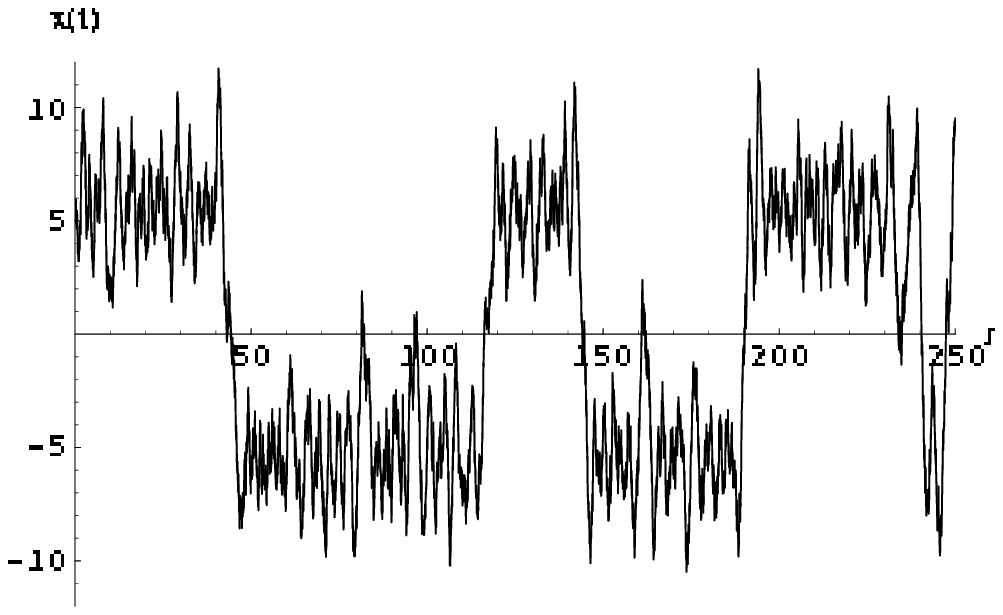,angle=0,width=12cm}}
\vskip 1cm
\caption{Long-time behavior of the mispricing for the uncertain 
weakly regulated market with individual reverting and collectively 
speculative agents, characterized by the phase portrait of Fig. 17. 
All market parameters are the same as in this figure 17. Initial 
conditions are: $x(0)=5$ and $y(0)=0$. The volatility parameter is 
$\sgm=2$. The trajectory jumps between the trapping regions 
corresponding to the points $\{0,0\}$, $\{-s,0\}$, and $\{ s,0\}$, 
and the trapping regions related to the cycles $C_{-s}$ and $C_s$. 
}
\label{fig:Fig.25}
\end{figure}

\end{document}